\documentclass[prd,aps,twocolumn,nofootinbib]{revtex4}
\usepackage[dvipdfm]{graphicx}
\usepackage{amssymb}
\usepackage{amsmath}
\usepackage{epsfig}
\newcommand{\ba}{\begin{eqnarray}}
\newcommand{\ea}{\end{eqnarray}}
\newcommand{\be}{\begin{equation}}
\newcommand{\ee}{\end{equation}}
\newcommand{\al}{\alpha}

\newcommand{\da}{\delta}
\newcommand{\la}{\lambda}
\newcommand{\ka}{\kappa}

\newcommand{\sa}{\sigma}
\newcommand{\en}{\epsilon}
\newcommand{\oa}{\omega}

\newcommand{\La}{\Lambda}

\newcommand{\cH}{{\cal H}}
\newcommand{\cF}{{\cal F}}

\newcommand{\cR}{{\cal R}}
\newcommand{\cO}{{\cal O}}

\newcommand{\cT}{{\cal T}}

\newcommand{\w}{\widetilde}
\newcommand{\x}{\star}

\newcommand{\p}{\partial}

\newcommand{\ra}{\rightarrow}
\newcommand{\Ra}{\Rightarrow}

\newcommand{\LF}{\left(}
\newcommand{\RF}{\right)}
\newcommand{\LT}{\left[}
\newcommand{\RT}{\right]}
\newcommand{\Ld}{\left.}

\newcommand{\kw}{\w{k}}
\newcommand{\tauw}{\w{\tau}}

\newcommand{\2}{\frac{1}{2}}
\newcommand{\3}{\frac{1}{3}}

\newcommand{\6}{\frac{1}{6}}

\newcommand{\mx}{\mbox}

\newcommand{\mand}{\mx{ and }}

\newcommand{\where}{\mx{ where }}
\newcommand{\with}{\mx{ with }}

\newcommand{\ie}{{\it i.e.\ }}

\newcommand{\vs}{\vspace{5mm}\\}
\newcommand{\non}{\nonumber\\}

\newcommand{\tb}{\bar{\tau}}
\newcommand{\rb}{\bar{\rho}}

\newcommand{\cHb}{\bar{\cH}}
\newcommand{\rhob}{\rho_{\bullet}}

\begin{document}

\preprint{IGPG-07/7-1}

\date{\today}

\title{Perturbations in Bouncing and Cyclic Models, a General Study}

\author{Tirthabir Biswas$^1$}
\author{Riley Mayes$^1$}
\author{Colleen Lattyak$^{1,2}$}
\affiliation{$^1$ Department of Physics, Loyola University,
New Orleans, LA 70118, USA}
\affiliation{$^2$ University of Illinois at Chicago (UIC), Chicago, IL 60607}

\pacs{98.80.Cq}

\begin{abstract}
Being able to reliably track perturbations across bounces and turnarounds in cyclic and bouncing cosmology lies at the heart of being able to compare the predictions of these models with the Cosmic Microwave Background observations. This has been a challenging task due to the unknown nature of the physics involved  during the bounce as well as the technical challenge of matching perturbations precisely between the expansion and contraction phases. In this paper, we will present general techniques (analytical and numerical) that can be applied to understand the physics of the fluctuations, especially those with ``long'' wavelengths, and  test its validity in some simple bouncing/cyclic toy models where the physics is well understood. We will then apply our techniques to more interesting cosmological models such as the bounce inflation and cyclic inflation.
\end{abstract}

\maketitle

\newcommand{\eq}[2]{\begin{equation}\label{#1}{#2}\end{equation}}

\section{Introduction}
It has long been appreciated that Standard Big Bang (BB) cosmology cannot
be the complete theory of the early universe since it is plagued by an
initial singularity~\cite{HP-singularity}. The singularity theorems have also
been extended to apply to scalar field-driven inflationary cosmology~\cite{Borde,Borde-Guth}. It is possible that the initial singularity is resolved
by quantum gravity where the space-time continuum description is replaced by a more fundamental description. Another possibility, however, is that an ``effective'' space-time description is possible even at Planckian energy densities and corrections
 to the gravitational Lagrangian and/or the description of matter stress-energy tensor  leads to a bouncing universe~\footnote{Another nonsingular way of resolving the BB singularity is to have an emergent universe~\cite{emergent1,emergent2} type of scenario where in the infinite past our universe approaches a Minkowski space-time, see also~\cite{emergent-cyclic}.}, for a recent discussion see~\cite{anupam-alex}. Traditionally, bouncing cosmology has been a popular alternative to the inflationary paradigm as during  the contraction phase it is relatively easy to arrange for the
Hubble radius $H^{-1}(t)$ to decrease faster than the physical wavelength of
fluctuations. Hence, it is conceivable that processes acting in the
contracting phase can lead to a non-inflationary mechanism for generating superHubble fluctuations that can explain the
origin of structure in the universe. Proposals which envisage such a mechanism include the
 Pre-Big-Bang~\cite{PreBigBang}, Ekpyrotic
\cite{ekpyrotic}, Matter bounce~\cite{Finelli,MatterBounce} and Hagedorn bounce~\cite{Nayeri,BBMS,BKM-hagedorn}.

More recently, it has been recognized that bounce and inflation need not be mutually exclusive leading to various Bounce-inflation scenarios~\cite{BM-bounce-inflation,Liu,Calcagni-BI,Piao-BI,Cai-Piao-BI} which opens up new ways to construct phenomenologically viable non-singular and geodesically complete models. For instance, in~\cite{BM-bounce-inflation} it was recognized that since a bounce necessarily implies a super-inflationary phase preceding ordinary inflation, the spectrum of modes exiting during the bounce would have a slight blue tilt as opposed to the conventional red tilt that is generated during the inflationary phase. This can potentially address the increase in power that has been observed in the low ($l\lesssim 40$) multipoles in CMB~\cite{WMAP9,Planck}.

Another class of bounce inflation scenarios have been considered in~\cite{BBKM,BKM-hagedorn} where the goal was to find a new mechanism that can yield a near-scale-invariant primordial spectrum during the contraction phase in the ``constant''  mode ($C_c$ mode) involving string thermodynamics. To be consistent with the observed amplitude of fluctuations in CMB, typically the fluctuating modes in these scenarios must exit the Hubble radius at very high temperatures $\sim 10^{15}$ GeV. This also means that the physical wavelengths of these fluctuations are very small and needs to be stretched  by about 60 efolds to be cosmologically relevant today. It then becomes imperative to invoke an inflationary phase following the contraction and bounce phases to make the models viable. However, an advantage of such a scenario  is that the inflationary regime need not be near-exponential, any accelerating phase, such as power-law inflation, would be sufficient. This may aid in building phenomenological models from fundamental theories such as string theory. It is worth pointing out that in other bouncing scenarios, such as in ekpyrotic models, the mechanism for producing scale-invariant perturbations proceeds via a different route: One typically produces a near scale-invariant power-spectrum in the ``growing''  mode ($G$ mode) during contraction. During the exit the amplitudes are ``small'' and the wavelengths are ``large'', but in the subsequent contraction phase the fluctuations are amplified to the observed $\sim 10^{-10}$ level. An additional mechanism is then introduced to transfer these fluctuations to the adiabatic constant mode.

A key aspect to assess the viability of all these different models that emerges from the above discussion is about how the fluctuations in the ``constant'' and ``growing'' modes in the contraction phase is transferred to the constant mode ($C_e$ mode) and ``decaying''  mode ($D$ mode) in the expanding phase. In particular, what matters is the spectrum in the constant mode of the expanding branch because that is what we see in CMB. For instance, for the success of the hagedorn bounce model~\cite{BKM-hagedorn}, we would want the near scale-invariant $C_c$-mode spectrum to be transferred to the $C_e$ mode. On the other hand, it would be advantageous for the ekpyrotic or matter bounce models if the $G$ mode spectrum could be transferred to the $C_e$ mode. This paper is primarily devoted to developing general strategies and techniques, both analytical and numerical, to study how the fluctuations are transmitted across the bounces.

Additionally we will also discuss evolution of perturbations in cyclic cosmological models which are extensions of the bouncing models that not only entail a bounce phase but also  turnarounds (transitions from contractions to expansions). In cyclic models such as the ekpyrotic cosmology, it is not particularly important to track the perturbations across the turnarounds as the fluctuations during the turnaround are too small to be relevant for the next cycle, and ``new'' perturbations are generated at each cycle. In other models~\cite{Cai-saridakis,Prokopec,cyclic-multiverse1,cyclic-multiverse2,Piao,Mulryne:2004va} however, perturbations that we see in CMB today are generated over many many short cycles and therefore it is important to be able to track the fluctuations through the  entire course of the cycle including the bounce and the turnaround. Therefore, in this paper our aim is to be able to track fluctuations across both turnarounds and bounces, and indeed over many cycles. We have succeeded in developing robust analytical methods to track the two usual modes of superHubble fluctuations, and have tested our methods using numerical computations. In the process we also found certain general trends that subHubble fluctuations seem to follow, but the robustness of these latter trends will need to be investigated further in future.

Before delving into the details of our framework, let us briefly discuss some of the challenges one faces in trying to understand the evolution of fluctuations in bouncing/cyclic cosmologies. While a vast body of literature already exists on the subject of perturbations around bouncing backgrounds, it is fair to say that we have relatively few general and conclusive results. The biggest hurdle to accomplishing this task in a comprehensive manner is simply
the fact that one doesn't yet have a theory of Quantum Gravity (QG), or even an effective description of it
that one can trust near the bounce which is expected to occur at the Planck scale~\footnote{
For progress in this direction see~\cite{Ashtekar,Bojowald,Shtanov,Freese,Baum,BMS,BGKM, palatini-bounce,cai-bounce}, for instance.}.
Often the bounce mechanism is mediated by "unphysical" and unstable ghost degrees of freedom,
and one of the reasons different groups have found different results
when it comes to tracking the perturbations could be because the underlying physical models
suffer from different inconsistencies/instabilities and therefore cannot completely be trusted.

To by pass our ignorance of the precise bounce mechanism often, the
evolution of the background and perturbations are modeled by a contracting GR phase
that is matched to an
expanding phase of Standard Big Bang cosmology through an
instantaneous and often singular transition along a space-like
hypersurface. For instance, one proposal has been to use the analog of the Israel
matching conditions (matching conditions~\cite{Israel} which describe
the merger of two solutions of the Einstein equations along a
time-like hypersurface).  These equations were discussed in
\cite{Hwang-Vishniac,Mukhanov-matching}. As pointed out by~\cite{martin1,peter1,peter2,martin2,martin3,durrer,Tsujikawa-flucts}, different matching prescriptions, however, can lead to different results thereby questioning the reliability of these prescriptions.

In this paper, our strategy will be to first focus on results that can be applied to a large class of  bouncing models. We will see that in a separate universe approach~\cite{separate-universe} (see also~\cite{Sasaki-Tanaka,Kodama-Hamazaki} it is possible to track two modes of superHubble fluctuations as long as we know the bouncing background, the precise knowledge of the new physics that makes a bounce possible will not be needed as long as  some basic symmetries are respected. We will test our results by looking at toy bouncing and cyclic models that can be constructed within the General Relativistic framework, and therefore does not suffer from any ambiguities coming from unknown new physics. We will then apply our framework to understand the evolution of superHubble perturbations in more realistic cosmological models, such as the emergent cyclic inflation~\cite{emergent-cyclic,AB-cyclic,BM-CI,BMS-cmb,BKM-exit,DB,essay} and bounce inflation~\cite{BM-bounce-inflation,Liu,Calcagni-BI,Piao-BI,Cai-Piao-BI}. In the process we will also be able to shed some light on how subHubble fluctuations evolve in cyclic cosmologies using numerical computations.

The paper is organized as follows: In section~\ref{sec:sup-hubble}, we will develop the analytical formalism to track super Hubble fluctuations exactly given the knowledge of the background evolution. In section~\ref{sec:numerical-test}, we will consider simple General Relativistic toy models which exhibits turnarounds and bounces and will obtain the evolution of perturbations numerically. We will then test the analytical results obtained in section~\ref{sec:sup-hubble} by comparing the numerical results  in the long wavelength limit. Next, in section~\ref{sec:bouncing}, we will  apply our formalism to study simple QG-mediated bouncing cosmologies and, in particular, demonstrate how one can obtain the transfer matrix relating the decaying and constant mode of the expanding branch to the constant and growing mode of the contracting branch. In section~\ref{sec:cyclic-inflation}, we will use our    results to investigate the stability and viability of the emergent-cyclic-inflation models. Subsequently, in section~\ref{sec:bounce-inflation} we will look at the perturbations in bounce inflation models which will also show us ways to go beyond the formalism developed as well as provide insights into the evolution of the subHubble fluctuations. We will conclude in section~\ref{sec:conclusions} by summarizing our main results and pointing out future research directions. Our appendix section~\ref{sec:appendix}, contains an illustrative example on how it is possible to derive the perturbation equations for superHubble fluctuations from the knowledge of the background evolution equations.
\section{Analytically Tracking Super-Hubble Fluctuations}\label{sec:sup-hubble}
Our aim in this section is to provide a general algorithm to analytically track certain superHubble modes. The key physics ingredient which allows us to provide such an algorithm is the realization that for superHubble fluctuations one can ignore the spatial fluctuations in comparison with the time variation. In other words, in the evolution equations of the superHubble fluctuations one only need to keep track of the time derivatives, but this is precisely what is done to arrive at the cosmological background solutions. Thus, it should be possible to obtain the evolution equations of the superHubble fluctuations even if we only  know the background evolution equation. In fact, if we already know the complete set of  background solutions, then we can obtain the superHubble modes directly from them, without ever having to even know what the superHubble perturbation equations are!

This  is really a variant of the separate universe approach. In the context of bouncing cosmology this approach was considered in~\cite{Wands}, but we provide a more general and comprehensive formalism with illustrative examples which also lets us verify the procedure with numerical precision. Our approach also provides several physical insights that could be helpful for model building purposes.
\subsection{Super Hubble Modes from Background Evolution}
In the {\it longitudinal gauge}, in the
absence of anisotropic stress,  the perturbed metric is given by
\be
ds^2 \, = \, a(\tau)^2 \bigl[ (1 + 2 \Phi) d \tau^2 -
(1 - 2 \Phi) d{\bf x}^2 \bigr] \, ,
\label{longitudinal}
\ee
where the relativistic potential $\Phi({\bf x}, \tau)$
is the field describing the (scalar) fluctuations. The idea is to be able to find the superHubble fluctuations in $\Phi$ from the knowledge of cosmological background solutions,
\be
ds^2 \, = \, a (\tb)^2\bigl[ - d \tb^2 + d{\bf x}^2 \bigr] \, ,
\label{cosmological}
\ee
that satisfy the appropriate cosmological equations. Since the conformal times associated with the metrics (\ref{longitudinal}) and (\ref{cosmological}) need not be the same, we have introduced a ``bar'' to distinguish the two. Suppose now that the set of all cosmological solutions are known and are characterized by the parameter set $\{\la_i\}$. These  parameters can be thought of as integration constants that arise when we solve the cosmological differential equations. Let us further suppose that we are interested in finding perturbations around a specific background solution chosen by convention to correspond to $\la_i=0,\ \forall\ i$. A small perturbation around the background metric, $a_b(\tb)=a(\tb,\{\la_i=0\})$, is now given by
\be
a(\tb)=a_b(\tb)[1+\sum_i\la_i f_i(\tb)]\where f_i\equiv \Ld{\ \p a\over a \p\la_i}\right|_{\la_i=0}
\label{cosmological}
\ee
Clearly, $f_i$'s are nothing but the different superHubble modes with $\la_i$'s the corresponding amplitudes. However, the fluctuations have been obtained in the so called synchronous gauge
\be
ds^2 \, = \, a (\tb)^2\LF1+2\psi\RF \bigl[ - d \tb^2 + d{\bf x}^2 \bigr] \, ,
\label{conformal}
\ee
\be
\with \psi=\sum_i\la_i f_i\ ,
\label{psi-f}
\ee
where for notational simplicity we have now dropped the subscript ``$b$''.
One needs to perform appropriate gauge transformations to obtain the more commonly discussed Newtonian potential, $\Phi$, in the longitudinal gauge (\ref{longitudinal}).

To find the relation between the two gauges we perform the following transformations:
$$ds^2 \, = \, a(\tau)^2 \bigl[ (1 + 2 \Phi) d \tau^2 -
(1 - 2 \Phi) d{\bf x}^2 \bigr]$$
$$\approx a(\tau)^2(1 - 2 \Phi) \LT (1 + 4 \Phi) d \tau^2 -
 d{\bf x}^2 \RT=a^2(1 - 2 \Phi) \bigl[  d \tb^2 -
 d{\bf x}^2 \bigr]$$
 where $\tau$ and $\tb$ are related as
 \be
 {d\tb\over d\tau}\approx 1+2\Phi
 \ee
Although the above metric is now in the same form as (\ref{conformal}), we see that there is an ambiguity in distinguishing the background scale factor from the perturbations. This has to be fixed by requiring form invariance of the scale factor. If $\xi$ relates the two time coordinates
\be
\tb=\tau+\xi\Ra \xi'=2\Phi
\label{xi-Phi}
\ee
then
\be
a(\tau)=a(\tb)-\xi a'(\tb)=a(\tb)(1-\xi\cH)
\ee
This gives us the relation between $\Phi$ and $\psi$:
\be
\psi=-\Phi-\xi\cH
\label{relation}
\ee
Thus, once we know $\psi$ from the cosmological solutions via (\ref{cosmological}) and  (\ref{psi-f}), we can  obtain the gauge parameter from the equation,
\be
\psi=-\2\xi'-\xi\cH\ ,
\label{psi-xi}
\ee
derived using (\ref{relation}) and (\ref{xi-Phi}). Subsequently, we can compute the Bardeen potential $\Phi$ from (\ref{xi-Phi}). In practice, it may not be possible to know all the cosmological solutions, but rather a subset of them. This can then be used to obtain a subset of the superHubble modes, as we will  explicitly illustrate in the next subsection.
\subsection{Super-hubble modes from time translation and scale-factor symmetry}
General Covariance essentially guarantees that the cosmological evolution equations preserve time translation symmetry. For flat space-time, we also have the so called scale-factor symmetry. Schematically the differential equations governing the cosmological evolutions must be of the form
\be
\cF_I\LF {\ \p^n a\over a \p t^n}, {\p^p \rho\over \p t^p}, {\p^p p\over \p t^p}\RF=0\with n>0\mand p,q\geq 0\ ,
\label{symmetry}
\ee
where $t$ is the proper time, which is related to the conformal time via $dt=a(t)d\tau$, as usual.
In particular, (\ref{symmetry}) implies that if $a(t)=a_b(t)$ is one particular solution of the evolution equation, then so is
\be
a(t)=(1+a_0)a_b(t-t_0)\ .
\ee
In conformal time the differential equations have a slightly different schematic form
\be
\cF_I\LF {1\over a}\LF{\ \p \over a \p \tau}\RF^na, \LF{\ \p \over a \p \tau}\RF^n\rho,\LF{\ \p \over a \p \tau}\RF^np\RF=0\ .
\ee
It is easy to check that for metrics written in the conformal gauge this means that the 2-parameter class of
 solutions is given by
\be
a(\tau)=(1+a_0)a_b[(1+a_0)(\tau-\tau_0)]
\ee
We emphasize that $a_b(\tau)$ is a specific function describing the background which does not contain any free parameters (integration constants).

Let us perturb the general solution around the specific background $a_b(\tau)$ to obtain the modes $f_i$'s discussed in the previous subsection:
\ba
a(\tau)&=&(1+a_0)a_b(\tau+a_0\tau-\tau_0)\non
&\approx& a_b(\tau)+a_0a_b(\tau)+(a_0\tau-\tau_0)a_b'(\tau)\non
&=&a_b(\tau)\LT1+a_0+(a_0\tau-\tau_0){a_b'(\tau)\over a_b(\tau)}\RT\ .
\ea
Thus, we have two superHubble modes in the synchronous gauge given by,
\ba
f_1(\tau)&=&1+\tau{a'(\tau)\over a(\tau)}=1+\tau\cH\\
\mand f_2(\tau)&=& -{a'(\tau)\over a(\tau)}=-\cH\ ,
\ea
associated with the amplitudes (integration constants) $a_0$ and $\tau_0$ respectively, where we have again dropped the subscript ``$b$'' for brevity. The corresponding Newtonian potential in the synchronous gauge reads
\be
\psi(\tau)=a_0\LT1+\tau\cH\RT-\tau_0\cH\ .
\ee

The next step is to compute the gauge parameter by solving (\ref{psi-xi}) which now reads
\be
a_0\LT1+\tau\cH\RT-\tau_0\cH+{1\over 2}\xi'+\xi\cH=0\ .
\ee
The above equation can be integrated to give us
\be
a^2(\tau)\xi+2\int_{\tau{\x}}^{\tau} d\w{\tau}\ a^2(\w{\tau})[a_0\LF1+\w{\tau}\cH(\w{\tau})\RF-\tau_0\cH(\w{\tau})]+{b_0\over M_p}=0\ ,
\ee
where $\tau_\x$ is any fixed time that can be chosen as convenient, and we have explicitly included the new dimensionless integration constant $b_0$ and introduced $M_p$,  the reduced Planck mass, for convenience. The integrals can be simplified as follows
$$\int^{\tau} d\w{\tau}\ a^2(\w{\tau})\LF1+\w{\tau}\cH(\w{\tau})\RF=\2\LT a^2\tau+\int^{\tau} d\w{\tau}\ a^2(\w{\tau})\RT$$
and
$$\int d\w{\tau}\ a^2 \cH=\int d\w{\tau}\ a {da\over d\w{\tau}}={1\over 2} a^2\ ,$$
where any further integration constants can be absorbed in the $b_0$ term.
The gauge parameter is thus given by
\be
\xi=-a_0\LT\tau+{g(\tau)\over a^2(\tau)}\RT-{b_0\over M_pa^2(\tau)}+\tau_0\ .
\ee
We have defined the function $g(\tau)$ as
\be
g(\tau)\equiv\int^{\tau}_{\tau{\x}} d\w{\tau}\ a^2(\w{\tau})
\label{g}
\ee

Finally, one can obtain $\Phi$ from the gauge parameter via (\ref{xi-Phi}):
\be
\Phi(\tau)=a_0(k)\LT{\cH g\over a^2}-1\RT+b_0(k)\LT{\cH\over M_pa^2}\RT
\label{suph-soln}
\ee
or in proper time:
\be
\Phi(t)=a_0(k)\LT{H g\over a}-1\RT+b_0(k)\LT{H\over M_pa}\RT
\label{suph}
\ee
with
\be
g(t)\equiv\int^{t}_{t{\x}} d\w{t}\ a(\w{t})
\ee
Note, that we have now included an explicit $k$-dependence in $a_0,b_0$ to emphasize the fact that the amplitude associated with the different superHubble modes will, in general, vary with the comoving wave number $k$. The $a_0$ mode is generally referred to as the constant  mode while the $b_0$ mode is referred to as the growing or decaying mode depending on whether the universe is contracting or expanding respectively.
\subsection{Possible Generalizations}
While most of this paper will be devoted to tracking the usual superHubble modes associated with the $a_0, b_0$ modes, the connection between the superHubble evolution and the background solutions discussed above is more general and can even be applied to cases where the scale-factor/time translation symmetry is not available, or when there are extra modes present. Here we will briefly discuss how our approach and results can be extended by looking at certain illustrative examples.
\subsubsection{The case of open or closed universe}
If the universe is spatially curved, \ie open or closed, then the scale-factor symmetry is no longer a good symmetry of the cosmological equation. For instance, consider the Hubble equation in an open/closed universe in the presence of an ideal fluid with energy density, $\rho$:
\be
H^2={1\over 3 M_p^2}\LF\rho-{K\over a^2}\RF
\ee
Due to the presence of the $a$ dependent curvature term, if $a(t)$ is a solution to the above equation, $(1+a_0)a(t)$ is no longer a solution. However, there is a way around this for an ideal fluid which satisfies the continuity equation
\be
\rho'+3\cH(1+w)\rho=0\ ,
\label{continuity}
\ee
 while, in conformal time, the Hubble equation in the presence of spatial curvature reads
\be
\cH^2={1\over 3 M_p^2}\LF a^2\rho-K\RF
\ee

Now, given a solution $a_b(\tau),\rho_b(\tau)$ of the above cosmological equations, it is easy to see that
\be
a(\tau)=(1+a_0)a(\tau-\tau_0) \mand \rho(\tau)=(1+a_0)^{-2}\rho_b(\tau-\tau_0)
\ee
also solves them~\footnote{This trick does not work for more general forms of matter such as scalar fields, or if the gravity side of the equation is modified.}. Proceeding in a manner similar to what was done in the previous subsection, one now arrives at a slightly different result for superHubble modes:
\be
\Phi(\tau)=a_0(k)\LT{2\cH g\over a^2}-1\RT+b_0(k)\LT{\cH\over M_pa^2}\RT
\label{suph-soln2}
\ee
We will see later how the above indeed reproduces the superHubble fluctuations in a numerical example. One may wonder that within GR it should be possible to employ the same argument for a flat universe as well to arrive at (\ref{suph-soln2}) instead of (\ref{suph-soln}); is there a discrepancy then? The answer obviously is no;  in GR for a flat universe, $\cH g/a^2$, is a constant making  (\ref{suph-soln2}) and (\ref{suph-soln}) equivalent to each other.
\subsubsection{Working with Evolution Equations}
There may be situations where one cannot find solutions associated with the time translation or scale factor symmetry or there could be additional modes present in the theory. How do we capture the superHubble evolution in these cases? Well, at least if the cosmological differential equations governing the background evolution is known, then it is possible to derive the perturbation equations that superHubble fluctuations must satisfy. In section~\ref{sec:appendix}, the appendix, we have illustrated this method for GR. Such a prescription may be a lot easier than obtaining the perturbation equations by perturbing the full gravitational equations. In some case, such as in Loop Quantum Gravity, the complete gravitational equations may not even be available, but our procedure can still  yield the equations relevant for superHubble fluctuations.

Once the perturbation equations are known, one can of course solve them, numerically if necessary. What often is of particular aid in this process is that in many phenomenological situations, far back in the past and far ahead in the future, the model reduces to GR.  The superHubble solutions in this case obviously are known and numerical solutions can then be used to connect the superHubble modes of the past with future via ``transfer matrix''. In a non singular geometry, as long one doesn't use pathological variables it should be possible to obtain these transfer matrices encoding the physics of the bounce and turnaround. We will illustrate this procedure using a simple bounce-inflation model  in section~\ref{sec:bounce-inflation}.
\section{Testing the SuperHubble Evolution Formula}\label{sec:numerical-test}
The main aim of this section is to look at GR based cosmological scenarios where the perturbations can be tracked precisely, at least numerically, and then check whether our superHubble formulae, (\ref{suph-soln}) and (\ref{suph-soln2}), holds. In the process, we will also be able to make some observations about the subHubble or short wavelength fluctuations.
\subsection{Monotonic Expansions \& Contractions}\label{sec:monotonic}
\setcounter{equation}{0}
We are going to first look at simple single fluid monotonically expanding or contracting FLRW models.  Starting with Einstein's equations
\be
G_{\mu}^{\nu}=T_{\mu}^{\nu}\ ,
\label{gen-GR}
\ee
 one can derive the usual General Relativistic perturbation equations for the Bardeen potential $\Phi$. For a perfect fluid (no anisotropic stress) the General Relativistic perturbation equations for $\Phi$ reads~\cite{Brandenberger}
\ba
 -k^2\Phi_k-3\cH(\Phi_k^{'}+\cH\Phi_k)  &=& {1\over 2M_p^2}a^2\da\rho_k\, \\
\Phi_k^{''} + 3\cH\Phi_k^{'} +
(2\cH^{'}+\cH^2)\Phi_k &=& {1\over 2M_p^2}a^2\da p_k\, .
\ea
The above equations can be combined to yield
\be
\Phi_k^{''} + 3(1+c_s^2)\cH\Phi_k^{'} + c_s^2 k^2\Phi_k +
[2\cH'+(1+3c_s^2)\cH^2]\Phi_k \, = \, 0 \, .
\label{gen-Phi}
\ee
where $c_s$ is the sound speed:
\ba
c_s^2\equiv {\da p\over\da\rho}\ .
\ea
For a single ideal fluid it is related to the equation of state parameter, $\oa$:
\be
c_s^2=\oa \where \oa\equiv {p\over\rho}\ .
\ee

(\ref{gen-Phi}) has the following super-Hubble limit:
\be
\Phi_k^{''} + 3(1+\oa)\cH\Phi_k^{'} +
[2\cH^{'}+(1+3\oa)\cH^2]\Phi_k \, = \, 0 \, .
\ee
For a single fluid evolution we further have
\be
a(\tau)= |M_p\tau|^q\ ; \cH={q\over \tau}\mand \cH'=-{q\over \tau^2}\ ,
\label{background}
\ee
with
\be
q\equiv {2\over 1+3\oa}\ .
\ee
Then the last term in (\ref{gen-Phi}) drops out and we are left with
\be
\Phi_k^{''} + {6(1+\oa)\over \tau(1+3\oa)}\Phi_k^{'}=0
\label{super-hubble}
\ee
In the above expression, $\tau>0$ represents an expanding phase while $\tau<0$ corresponds to a contracting universe.
(\ref{super-hubble}) has two well-known solutions, the constant mode and the decaying/growing mode:
\be
\Phi=C+{D\over |M_p\tau|^{2q+1}}
\label{phi-soln}
\ee

Let us now verify that (\ref{suph-soln}) indeed reproduces the above two modes. Let's specialize to the expanding universe, the calculations for the contracting universe follow similarly. For (\ref{background}) the function $g$ is given by
\be
g=\int d\tau a^{2}(\tau)= M_p^{-1}\LF{1+3\oa\over 5+3\oa}\RF(M_p\tau)^{2q+1}
\ee
(\ref{suph-soln}) then yields
\be
\Phi(\tau)=-{3a_0(1+\oa)\over 5+3\oa}+{2b_0\over (1+3\oa)(M_p\tau)^{2q+1}}\ ,
\ee
which is indeed of the form (\ref{phi-soln}).
\subsection{Perturbations Around a Turnaround}\label{sec:turnaround}
\subsubsection{The toy model}
Tracking fluctuations around turnarounds is easier than during bounce primarily because in most cyclic models turnarounds occur at energy densities much below the Planck scale where GR is assumed to hold. Thus no ``new physics'' is required to attain a turnaround and the evolution equations are well known. Taking inspiration from the cyclic inflation model~\cite{essay,BM-CI,BMS-cmb,BKM-exit,BKM-cyclic}, we will here consider a turnaround mediated by a negative cosmological constant in the presence of radiation. Accordingly, the Hubble  equation can be written as
\be
H^2={\rho\over 3M_p^2}\with \rho=\La(a^{-4}-1)\ ,
\label{hubble}
\ee
We have chosen our conventions such that the turnaround occurs  at $a=1$. $\La$ is a free parameter and governs the time scale of the turnaround. For numerical simulations it becomes imperative that we work with the Ray Chaudhury (RC) equation~\footnote{Numerically,  the Hubble equation can only be solved separately for the expanding and contracting phases as one needs to take its  square root before it can be implemented in a code. Thus it is not possible to track the evolution of the background or the perturbation through bounces and turnarounds using the Hubble equation. The Hubble equation, however, acts as a constraint determining the initial condition for $a'$.}:
\be
{\ddot{a}\over a}=-{1\over 6M_p^2}(\rho+3p)
\ee
where
\be
p=p_r+\La=\La\LF {a^{-4}\over 3}+1 \RF
\label{pressure}
\ee
Also, in order to compare our numerical results with the analytical expressions of (\ref{suph-soln}) we have to work with conformal time. The  Hubble and RC equations read
\be
\cH^2={a^2\rho\over 3M_p^2}\ \mand\
{a''\over a}={a^2\over 6M_p^2}(\rho-3p)\ .
\label{rc}
\ee

Let us next look at the perturbation equation, (\ref{gen-Phi}). Since the cosmological constant does not contribute to any fluctuations, the fluctuations in pressure and energy density only comes from radiation, and therefore we have
\be
c_s^2=\3\ .
\ee
(\ref{gen-Phi}) then can be simplified using the Hubble (\ref{hubble}) and RC (\ref{rc}) equations to yield
\be
\Phi_k^{''} + 4\cH\Phi_k^{'} + \3 k^2\Phi_k -{4\La\over 3M_p^2}a^2\Phi_k \, = \, 0 \, .
\ee
\subsubsection{Numerical Analysis}
It is rather useful to work in units of the time-scale associated with the bounce:
\be
\cT\equiv \sqrt{M_p^2\over \La}
\ee
Accordingly, the relevant dimensionless variables are
\be
\w{\tau}\equiv {\tau\over \cT}\ ;\ \w{\cH}=\cT\cH={1\over a}{da\over d\w{\tau}}\mand \w{k}= k\cT
\ee
The evolution equation now becomes
\be
\Phi_k^{''} + 4\w{\cH}\Phi_k^{'} + \3 \kw^2\Phi_k -{4\over 3}a^2\Phi_k \, = \, 0 \, ,
\ee
where the prime now denotes differentiation with respect to the $\tauw$ variable. The above equation can be solved in conjunction with the acceleration equation (\ref{rc}). Note, that the initial condition for $a'$ can be determined from the rescaled Hubble equation as follows:
\be
{da\over d\tauw}=+\sqrt{a^4\cT^2\rho\over 3M_p^2}=\sqrt{1-a^{4}\over 3}
\ee
\begin{figure}[htbp]
\begin{center}
\includegraphics[width=0.40\textwidth,angle=0]{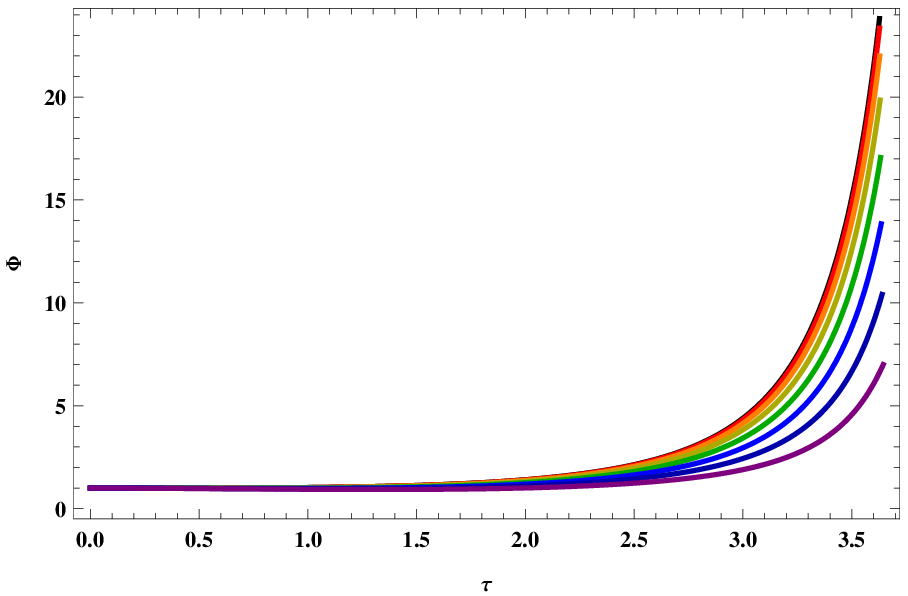}
\end{center}
\caption{Numerical evolution of $\Phi$ for $k\cT=$ 0.2, 0.4, 0.6, 0.8, 1.0, 1.2, 1.4 (red, orange, ochre, green, blue, indigo, purple) during a turnaround. The black curve corresponds to the analytical result for the superHubble evolution. Evidently, the colored curves approach the black as $k\ra 0$. The initial conditions for the above graph corresponds to $a(0)=a_0=0.1$ and $\Phi'(0)=0$.
  \label{fig:turnaround}}
\end{figure}

Fig.~\ref{fig:turnaround} depicts the evolutions of $\Phi$ for different values of the comoving wave number $\w{k}=k\cT$. The red curve denotes the analytical result (\ref{suph-soln}) and as is evident, the blue curves approach it in the $k\ra 0$ limit. Firstly, this  corroborates the validity of our separate universe approach to track the perturbations during transitions between expansion and  contraction, and in particular the formula (\ref{suph-soln}).

Secondly, it supports the view point~\cite{Kinney,BMS-cmb} that the Hubble parameter is not always the most appropriate comparative ``yard-stick'' in determining what constitutes a  superHubble fluctuation. For instance, exactly at the turnaround point, $H=0$, and therefore technically, all modes are subHubble! Intuitively however, we know that such a criteria is not physically sensible; for instance, for sufficiently long wavelengths the time their wavelengths  remain smaller than $H^{-1}$ is too short to even establish a causal connection over the wavelength itself. Thus the ``subHubble'' phase should not have any effect on evolution for these modes, and the modes should essentially follow the superHubble trajectory. This is exactly what our numerical simulations indicate: Modes with sufficiently small $k\ll \cT^{-1}$ follows the analytical superHubble trajectory (\ref{suph-soln}). In other words, the correct time scale that one should compare the physical wavelengths of fluctuations to determine whether the mode remains ``superHubble'' is the turnaround time scale, given in this model by $\cT$. We will see that the similar arguments also hold for bouncing cosmologies. This  is an important result in the context of cyclic/bouncing cosmologies as it greatly simplifies tracking long wavelength fluctuations.
\subsubsection{SubHubble Fluctuations}
\begin{figure}[htbp]
\begin{center}
\includegraphics[width=0.40\textwidth,angle=0]{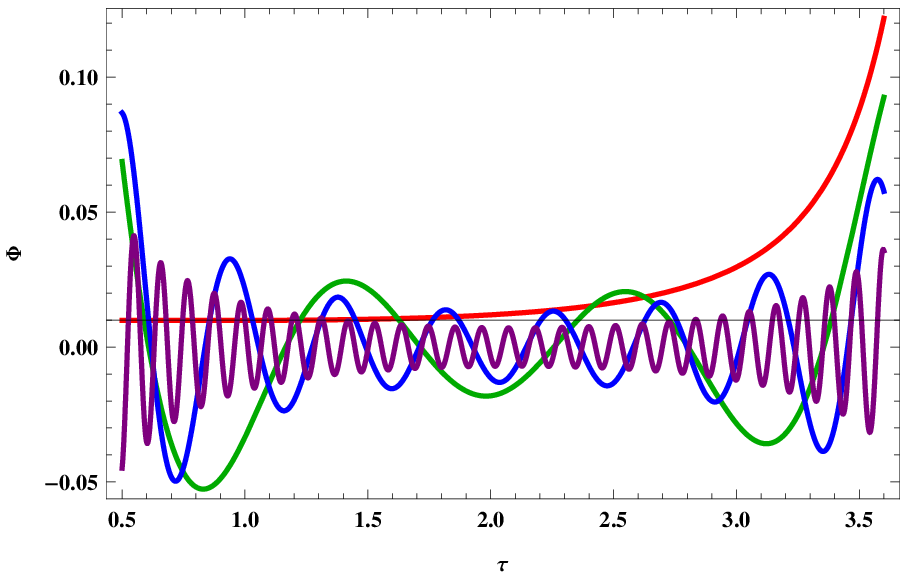}
\includegraphics[width=0.40\textwidth,angle=0]{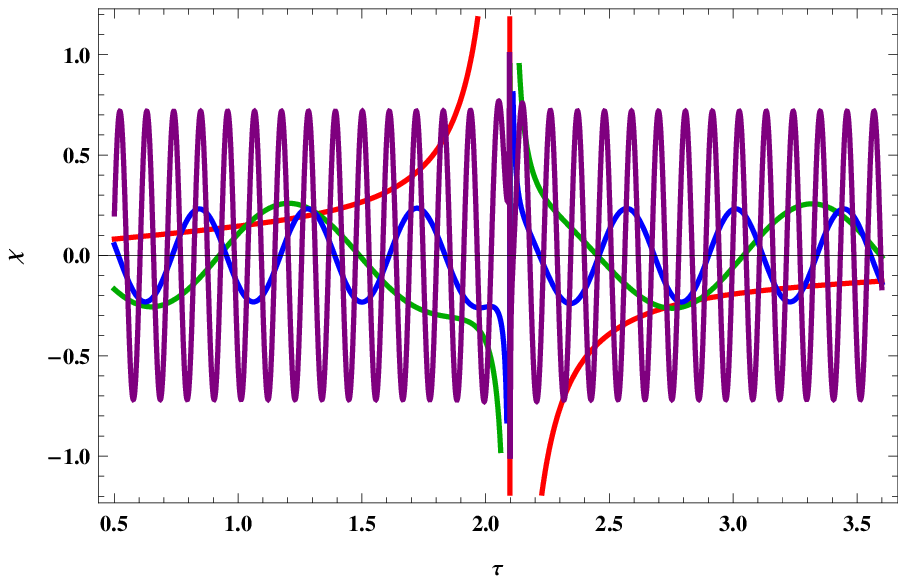}
\end{center}
\caption{Top: Numerical evolution of $\Phi$ for subHubble modes: $k\cT=$ 1, 10, 25, 100 (red, green, blue, purple) during a turnaround. The black curve corresponds to the analytical result for the superHubble evolution. Bottom: Numerical evolution of $\chi$ for the same modes. Evidently, the amplitude of fluctuations in $\chi$ stays a constant for the subHubble modes except at the bounce point where it is ill-defined. The initial conditions for the above graphs correspond to $a(0)=0.1$ and $\Phi'(0)=0$.
  \label{fig:turnaround-sub}}
\end{figure}

What can we say about the evolution of the subHubble fluctuations? Unlike the superHubble fluctuations where the spatial fluctuations did not matter  because their wavelengths were too large as compared to the cosmological evolution,  the same cannot obviously be argued for subHubble fluctuations. Moreover, often QG theories involve higher derivatives, which would necessitate keeping track of higher powers of $k$ in the evolution equations. The situation is complicated and model dependent, and to our knowledge no general approach to tracking subHubble fluctuations have been discussed in the previous literature. In Fig.~\ref{fig:turnaround-sub} we have plotted the evolution of $\Phi$ for different subHubble wavelengths and they oscillate with time-varying amplitudes. Can we find any general trend?

It is known that for monotonically expanding or contracting GR models with a single fluid or a scalar field although, the amplitude of $\Phi$ oscillations fluctuate, the oscillations in the so-called ``canonical'' variable, $\chi$, have constant amplitude~\cite{Brandenberger}. This is because the subHubble wavelengths are so short as compared to the cosmological time scale that one can find a canonical variable where the cosmological dynamics almost freezes and  the corresponding action resembles that of a free-field in Minkowski space-time. What happens to $\chi$ during a turnaround?

For a cosmology dominated by ideal fluids the canonical variable is defined via~\cite{Brandenberger}
\ba
\chi&\equiv& z\xi,\label{chi-defn}\\
\with\xi&=&\Phi+{2\over 3(1+p/\rho)}\LF{\Phi'\over \cH}+\Phi\RF\ ,\\
\mand z&=& a\sqrt{1-{\cH'\over \cH^2}}\ .
\ea
Clearly, $\chi$ is not a good variable exactly at the turnaround point since $z$ diverges at this point. The same happens during a bounce as we will see in section~\ref{sec:bounce-inflation}. Sometimes the divergence of $\chi$ is taken as evidence that a bounce is inherently unstable, but as seen in Fig.~\ref{fig:turnaround} and~\ref{fig:turnaround-sub}, the metric is completely well-defined and the perturbations can remain well under control. In section~\ref{sec:bounce-inflation}, we will see the same phenomenon repeating during a bounce. This is a clear demonstration that bounces and turnarounds are not inherently unstable and the cosmology can remain true to the FLRW assumption of a homogeneous universe.

What is perhaps more interesting is the observation that the amplitude of subHubble fluctuations in $\chi$ does not change during the course of the bounce, see Fig.~\ref{fig:turnaround-sub} (bottom). Thus our intuition based on monotonic cosmologies in GR about $\chi$ oscillations  seems to continue to hold. If such a result could be proven for more general situations, it would be a tremendous step in our understanding of perturbative evolutions, and in section~\ref{sec:bounce-inflation}, we will argue why there is good reason to be optimistic about this trend.
\subsection{Perturbations Around Bounce}\label{sec:brane-gas}
\subsubsection{Toy model with gas of branes}
In this subsection, the plan is to study how perturbations evolve around a bounce. For the purpose of illustration, we are again going to consider a toy model that can produce a bounce  within the confines of the General Relativity. We are going to look at a closed universe model containing a perfect fluid satisfying $-1<\oa<-1/3$. In this case the background cosmology gives rise to a bounce inflation scenario where a fast contraction is followed by an accelerated expansion mediated via a nonsingular bounce. Inflation is of a power-law type which approaches the exponential case in the limit $\oa\ra -1$. A particular example of an ideal fluid with such negative equation of state parameter is a gas of co-dimension 2 branes (or domain walls), which has an equation of state $\oa=-2/3$. It is also possible to mimic such equation of state parameters with a gas of strings appropriately coupled to scalars such as the dilaton, see for instance~\cite{BBEM}. We are however not really interested in the phenomenology of such models; our purpose is to test our formalism for tracking superHubble fluctuations in bouncing backgrounds where explicit numerical calculations can be performed.

With this in mind, let us proceed to obtain the differential equations that govern the evolution of the background and the perturbations. The total ``effective'' energy density and pressure  that enters  the Hubble equation (\ref{hubble}) and the RC equation (\ref{rc}) can be written as
\ba
\rho&=&\rho_f-{\rho_{\bullet}\over a^2}=\rho_{\bullet}(a^{-3(1+\oa)}-a^{-2})\\
p&=&p_{f}+\3 {\rho_{\bullet}\over a^2}=\rho_{\bullet}\LF\oa a^{-3(1+\oa)}+\3 a^{-2}\RF\ ,
\ea
where  we have chosen a convention such that the bounce occurs at $a=1$.

\begin{figure}[htbp]
\begin{center}
\includegraphics[width=0.40\textwidth,angle=0]{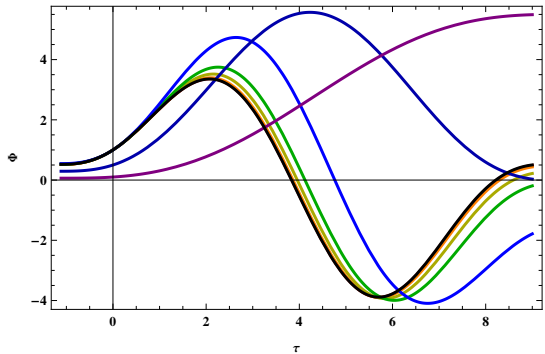}
\end{center}
\caption{Numerical evolution of $\Phi$ for $k\tau_{\bullet}=$ 0.05, 0.1, 0.2, 0.3, 0.5, 0.8, 1.0 (red, orange, ochre, green, blue, indigo, purple) during the bounce. The black curve corresponds to the analytical result for the superHubble evolution. Evidently, the colored curves approach the black as $k\ra 0$. The initial conditions for the above graph corresponds to $a(0)=10$ and $\Phi'(0)=0$.
  \label{fig:bounce}}
\end{figure}

The perturbation equation for the Bardeen potential needs to be modified to account for the curvature of the universe and according to our conventions reads~\cite{Brandenberger}
\ba
& &\Phi_k^{''} + 3(1+c_s^2)\cH\Phi_k^{'} +  c_s^2k^2\Phi_k
\non
 &+&\LT 2\cH'+(1+3c_s^2)\LF \cH^2-\3\RF\RT\Phi_k= 0 \, .
\label{closed-Phi}
\ea
The fact that it is negative simply indicates the instability of the system at short distances, but here we are only interested in the long wavelength fluctuations. In Fig.~\ref{fig:bounce}, we have illustrated the behavior of $\Phi$ for different values of $k$ for the domain wall case corresponding to $\oa=-2/3$.

In Fig.~\ref{fig:bounce}, we have plotted $\Phi_k$ versus the re-scaled conformal time $\tau\ra \tau/\tau_{\bullet}$, where $\tau_{\bullet}={M_P^2/\rho_{\bullet}}$ is the bounce time scale. Again, we see that as $k\ra 0$, the evolution approaches the analytically derived  superHubble evolution (\ref{suph-soln2}).
\section{Application I: Transfer Matrix around Bounce}\label{sec:bouncing}
In the previous sections we have derived  analytical expressions for the time evolution of the two superHubble modes, and verified their validity in  cosmological models involving monotonic expansions or contractions as well as when there is a transition between the two phases. We performed these checks in GR based models where  an exact numerical analysis could be performed. In the next couple of chapters our goal is to apply our derived superHubble formula (\ref{suph-soln}) to understand the evolution of perturbations for certain classes of nonsingular cosmological models. On one hand, this will serve as an illustration of how one can apply our formalism and results to different cosmological models, and on the other, it will shed insights into these specific scenarios  that we hope will be useful in building viable and consistent cosmologies.

In cosmology, we are most often interested in calculating the ``primordial'' spectrum of fluctuations that seeds the temperature anisotropies observed in the CMB. The usual algorithm is the following. One starts with a choice about the initial conditions (IC's) for the fluctuations in the subHubble phase. Traditionally, one uses the Bunch-Davis vacuum to obtain the IC's. Although other vacuum choices~\cite{nonBD} as well as statistical thermal initial conditions have  been explored~\cite{param,Cai-thermal,BBKM}, fortunately, the underlying algorithm for calculating the primordial spectrum does not really depend on the IC prescription. Typically, the fluctuations are tracked as they pass from the subHubble to the superHubble phase at which point the perturbations are ``expected to freeze''. In this paper, primarily we are concerned with the superHubble evolution that ensues. Do perturbations really freeze, even when there are cosmological transitions such as bounces and turnarounds? How is the ``initial'' spectrum that is generated while Hubble crossing (possibly during a contraction phase) related to the ``primordial spectrum'' in the expanding branch that is suppose to seed our CMB fluctuations?

Of course, we already have an analytical formula for the superHubble fluctuations, but to address these questions it is often more convenient to obtain a ``transfer matrix'' that relates how the amplitudes of fluctuation of the decaying and constant mode of the expansion phase is related to the growing and constant mode of the contraction phase when the expansion and contractions are connected via a nonsingular bounce or turnaround. In this section, we are going to illustrate how our formalism can be used to answer these questions within the framework of a simple bouncing cosmology.
\subsection{A Simple Bounce Cosmology}
Let us consider a simple but consistent example of a  bouncing universe dominated by a single fluid with $\oa>0$, so  that, unlike the brane-gas scenario discussed in section~\ref{sec:brane-gas}, the underlying thermodynamic system is stable. The price we have to pay is that we can no longer stay within the framework of GR as various no-go theorems exists that precludes bounce in GR with regular ($\oa>0$) matter~\cite{HP-singularity,Borde,Borde-Guth}. Thus, we envision having a contracting GR phase that lasts approximately till Planckian energy density is reached. This is  followed by a QG bounce with a Planckian cosmological time scale that eventually leads to a GR expanding branch once the energy density has diluted to sub-Planckian values. Now, during such an evolution, there are going to be some modes which are so short that even at the bounce point their wavelengths are shorter than the cosmological time scale, $M_p^{-1}\sim 10^{-42}$s. These modes remain subHubble throughout the course of the evolution and we will not discuss their evolution any further. All other modes exit the Hubble radius some time in the contracting GR phase and remain superHubble during the course of the bounce. Let us remind ourselves  that since the time scale of the bounce is $\sim M_p^{-1}$,  all the modes whose physical wavelength is larger than the Planck length near the bounce behave as super-Hubble modes.
\begin{figure}[htbp]
\begin{center}
\includegraphics[width=0.40\textwidth,angle=0]{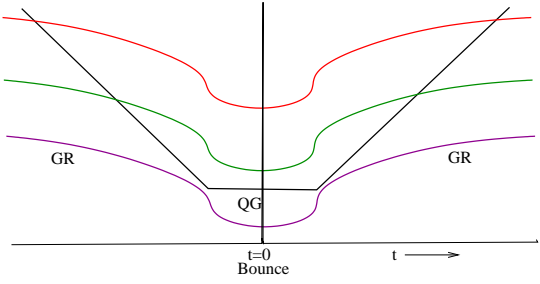}
\end{center}
\caption{Schemtic Hubble diagram for an $\oa>0$ universe. Black curve represents the cosmological time scale, while the colored curves correspond to the physical wavelengths for specific comoving modes. All the modes start out sub Hubble in the GR phase. Most of them exit the Hubble radius during the contracting phase, and re-enter the Hubble radius in the expanding branch. Some very short modes (purple curve) can always remain sub Hubble.
  \label{fig:hubdiagram1}}
\end{figure}

Now, as discussed in section~\ref{sec:monotonic}, in the GR regime the super-Hubble fluctuations are known to be of the form (\ref{phi-soln})
\be
\Phi(\tau)=C_{\pm}\pm {D_{\pm}\over |M_p\tau|^{2q+1}}\ .
\label{GRsup}
\ee
The $-$ and $+$ indices referring to the super-Hubble phases in the ``past'' and the ``future'' branch respectively. To gain insights into the evolution we want to find the transfer matrix relating $C_+,D_+$ with $C_-,D_-$. For instance, if the matrix is found to be diagonal then we can conclude that there is no mixing between the modes; the growing mode during contraction goes over to the decaying mode during expansion, while the constant modes on the two sides of the bounce map to each other. Another important issue has to do with whether there is any amplification/reduction in the amplitude of fluctuations as we go through the bounce. As a final point, we should also check whether the fluctuations can remain small during the entire bounce phase for us to justify using perturbation theory.

With these goals in mind, let us begin by choosing a coordinate system such that $\tau=0$ corresponds to the bounce. Then, in order to obtain the transfer matrix, all we have to do is to  expand the  solution (\ref{suph-soln}) that is valid in the entire super Hubble phase as a series in $(M_p\tau)^{-1}$ for large times~\footnote{This is assuming that non-GR effects appear at Planck scale which is then also identified as the bounce scale. If some lower mass scale effects (such as may be from string theory) are present then one has to look at appropriately larger times. This however doesn't change the basic algorithm proposed here.}. The  leading and next to leading order terms should then correspond to the two super Hubble modes in the GR phase. This way one can determine $C_{\pm},D_{\pm}$ in terms of $b_0,a_0$, and subsequently the transfer matrix.

Now, we know that as long as $|\tau|\gg M_p^{-1}$   we can trust the GR evolution and accordingly
\be
a_E(\tau)\approx \en_{\pm}|M_p\tau|^q\ ,
\label{GR-scalefactor}
\ee
where the $+,-$ indices again indicated the regions, $ M_p\tau\gg 1$ and $M_p\tau\ll -1$ respectively. For a symmetric bounce $\en_+=\en_-$, but not otherwise.

Next, let us define
\be
g\equiv \int_0^{\tau}a^2(\tb)d \tb
\ee
While evaluating the integral for positive $\tau$, we can break up the above integral into two parts: $[0,\tau_{\bullet}]$ and $[\tau_{\bullet},\infty)$, with $\tau_{\bullet}\gg M_p^{-1}$. In this case $g(\tau)$ can be evaluated as follows: For $\tau>\tau_{\bullet}$
\ba
g(\tau)&=&\int_0^{\tau_{\bullet}} d\bar{\tau}\ a^2(\bar{\tau})+\int_{\tau_{\bullet}}^{\tau} d\bar{\tau}\ a^2(\bar{\tau})\non
&\approx& \int_0^{\tau_{\bullet}} d\bar{\tau}\ a^2(\bar{\tau})+\int_{\tau_{\bullet}}^{\tau} d\bar{\tau}\  a_E^2(\bar{\tau})\non
&=& \int_0^{\tau_{\bullet}} d\bar{\tau}\ [a^2(\bar{\tau})-a_E^2(\bar{\tau})]+\int_{0}^{\tau} d\bar{\tau}\  a_G^2(\bar{\tau})\non
&\equiv & g_+ +\int_0^{\tau} d\bar{\tau}\ a_E^2(\bar{\tau})
\ea
The second term can be integrated straightforwardly to give us
\be
g(\tau)= g_++{\en_+(M_p\tau)^{2q+1}\over M_p(2q+1)}\mx{ for }\tau\gtrsim\tau_{\bullet}\ .
\ee
We note that since for large $\tau$, $a(\tau)\ra a_E(\tau)$, and $g_+$ tends to a constant as one increases $\tau_{\bullet}$, as it must for the approximation to be consistent. For $\tau\lesssim-\tau_{\bullet}$ one can perform a similar analysis by breaking the integral as
$$\int_0^{\tau}=-\LT\int^0_{-\tau_{\bullet}}+\int^{-\tau_{\bullet}}_{\tau}\RT$$
leading to
\be
g(\tau)= -\LT g_-+{\en_-(-M_p\tau)^{2q+1}\over M_p(2q+1)}\RT\mx{ for }\tau\lesssim-\tau_{\bullet}\ ,
\ee
where
\be
g_{\pm}\equiv \pm\int_0^{\pm\tau_{\bullet}} d\bar{\tau}\ [a^2(\bar{\tau})-a_E^2(\bar{\tau})]
\ee
Again, for a symmetric scale factor we should have $g_+=g_-$. To summarize, quite generically we have
\ba
a(\tau)&\ra& \en_{\pm}|M_p\tau|^q\label{a-form}\\
\cH &\ra& {q\over \tau}\\
g &\ra& \pm\LT g_{\pm}+{\en_{\pm}|M_p\tau|^{2q+1}\over M_p(2q+1)}\RT\ ,\label{g-form}
\ea

We are now in a position to evaluate $\Phi$ for large times:
\be
\Phi_{\pm}(\tau)=a_0\LF{\en_{\pm}q\over 2q+1}-1\RF+{q(a_0 M_pg_+\pm b_0)\over |M_p\tau|^{2q+1}}\ .
\label{super-expansion}
\ee
We now explicitly see that for $|\tau|>\tau_{\bullet}$, the first and the second terms in (\ref{super-expansion})  correspond to the constant mode and  the growing/decaying  mode respectively. By  matching the above asymptotic expansion with the GR super-Hubble solutions (\ref{GRsup}), we can  find the transfer matrix that we are after. We have
\ba
C_{\pm}&=&a_0\LF{\en_{\pm}q\over 2q+1}-1\RF\\
D_{\pm}&=&q(a_0 M_p g_{\pm}\pm b_0)
\ea
One can rewrite the above equations in a transfer matrix language:
\be
\LF
\begin{array}{c}
C_+\\
D_+
\end{array}
\RF=
\LF
\begin{array}{cc}
{q(\en_+-2)-1\over q(\en_--2)-1}  &0 \\
-{q(2q+1)M_p(g_++g_-)\over q(\en_--2)-1}&-1
\end{array}
\RF
\LF
\begin{array}{c}
C_-\\
D_-
\end{array}
\RF
\ee

Several comments are now in order: Firstly, it is now clear that while the modes do mix, the constant mode in the expanding branch does not get any contribution from the growing mode in the contraction phase. This is a crucial result for the purpose of model building in bouncing and cyclic models. For instance, it corroborates the result~\cite{Wands}, that in a single fluid scenario, at least, the scale-invariant perturbations of the growing mode in ekpyrotic cosmology cannot be transferred to the constant mode in the expanding branch that is relevant for CMB. Secondly, we see that for a symmetric bounce $C_+=C_-$, \ie, there is no amplification in the constant mode. However, if the bounce is not symmetric, \ie $\en_+\neq \en_-$, the amplitude may change. In a single fluid universe it is hard to imagine how one can realize an asymmetric bounce, but our results strongly suggest that amplification may be possible in a multi-component asymmetric bounce.

Finally, let us discuss the general behavior of the superHubble perturbations near the bounce. Let us look at the $b_0$ mode first. We see that in the GR contracting phase the mode grows very fast $\sim |M_p\tau|^{-(2q+1)}$. However, as the universe enters the QG phase around $\tau\sim M_p^{-1}$, $\cH\ra 0$, and the scale factor reaches a minimum, so that approximately one expects its largest value to be given by $\cH/M_p a^2\sim\cO(1)$. For the validity of the perturbative analysis then $b_0\ll 1$, but importantly, the perturbations do not grow unboundedly during the bounce.

What about the $a_0$ mode? Clearly it begins as a constant in the infinite past. According to our conventions, while $|\cH|/a^2$ keeps growing in the contracting phase $|g|$ decreases, so that the overall growth of the $a_0$ mode is slightly weaker than the $b_0$ mode. Exactly at the bounce, again $g=\cH=0$ while $a$ is at its minimum (but finite) value. Thus again, the perturbations remain bounded and attains its maximum value of $|g\cH/a^2|\sim \cO(1)$ around $\tau\sim M_p^{-1}$. To summarize, as long as $a_0,b_0\ll 1$, although both the superHubble modes grow towards the bounce, they can remain small enough for perturbation theory to remain valid, and the bounce scenario to be consistent!
\subsection{Illustrative Examples}
Our formalism is particularly useful in situations where the evolution equations for the perturbations are either not known or too complicated.
Let us therefore illustrate our algorithm in one such example; bouncing cosmology arising in Loop Quantum Cosmology where although the background equations have been studied in great depth, a complete understanding of the perturbations is still lacking. We will start with the modified Hubble equation that appears in LQC~\cite{Ashtekar,Bojowald}:
\be
H^2={1\over 3M_p^2}\LF \rho-{\rho^2\over \rho_{\bullet}}\RF
\label{lqc}
\ee
where $\rho_{\bullet}\sim M_p^4$. Thus at low energy densities, $\rho\ll M_p^4$, we recover the usual Hubble equation.

For an ideal gas, this gives us
$$H^2={\rho_{\bullet}a^{-2n}\over 3M_p^2}(a^{n}-1)\mx{ with }n\equiv 3(1+\oa)\ ,$$
which can be recast into
\be
{da^n\over dt}=n\sqrt{\rho_{\bullet}(a^n-1)\over 3M_p^2}
\ee
A straight forward integration yields
\be
 a=\LF1+{n^2t^2\rho_{\bullet}\over12M_p^2}\RF^{1/n}
\label{lqc-soln}
\ee
For large values of $t$ we clearly recover the GR limit.

We are now going to expand the LQC perturbed solution  in the small variable $x\equiv 1/|M_pt|$ to obtain the superHubble limit in the GR phase. Specifically, for the background solution (\ref{lqc-soln}), one finds
\ba
a(t)&=&\al^{1\over n} x^{-2\over n}\LF1+ {x^2\over \al}\RF^{1\over n}\approx\al^{1\over n} x^{-2\over n}\LF1+ {x^2\over n\al}\RF \nonumber\\
H&=& \pm{2\al M_p x\over n(1+x^2/\al)}\approx {2\over n}\al M_p x\LT1- {x^2\over \al}\RT \nonumber\\
g&=&{\pm1\over M_px} \, _2F_1\left(\frac{1}{2},-\frac{1}{n};\frac{3}{2};-{\alpha\over x^2}  \right)
\ea
where
\be
\al\equiv {n^2\rho_{\bullet}\over   12M_p^4}\ ,
\ee
and $_2F_1$ is a hypergeometric function with the limiting expression
\ba
 _2F_1\left(\frac{1}{2},-\frac{1}{n};\frac{3}{2};-{\alpha\over x^2}  \right)&=&x\LT\frac{\sqrt{\frac{\pi }{\al}} \Gamma \left(-\frac{1}{2}-\frac{1}{n}\right)}{2 \Gamma \left(-\frac{1}{n}\right)}+\cO(x^2)\RT\non
 &+&\LF{\al\over x^2}\RF^{1\over n}\LT\frac{n}{n+2}+\cO(x^2)\RT\ ,\non
\ea
as $x\ra 0$ or $M_p\ra \infty$. Thus the dominant terms in the large $t$ limit are given by
\be
g={\pm1\over M_p} \LT\frac{\sqrt{\frac{\pi }{\al}} \Gamma \left(-\frac{1}{2}-\frac{1}{n}\right)}{2 \Gamma \left(-\frac{1}{n}\right)}+ x^{-{n+2\over n}}\al^{1\over n}\LF\frac{n}{n+2}\RF\RT\ .
\ee

To compare our results with previous analysis we will need to know how the conformal time  is related to physical time $t$ in this model. We choose
\be
t = \int_0^\tau a(\tb) d\tb\ ,
\ee
which for the GR limit gives us
\ba
& &|M_p\tau|=[(1+q)|M_pt|]^{1\over 1+q}\\
&\Ra& (1+q)^{1+2q\over 1+q} |M_p t|^{{n+2\over n}}=|M_p\tau|^{1+2q}
\ea
It is easy to check that $a,g$ indeed reduces precisely to the generic form enumerated in (\ref{a-form}) and (\ref{g-form}) with
\ba
g_+=g_-&=& \frac{\sqrt{\frac{\pi }{\al}} \Gamma \left(-\frac{1}{2}-\frac{1}{n}\right)}{2 M_p \Gamma \left(-\frac{1}{n}\right)}\\
\mand\en_+=\en_-&=& \al^{1\over n}(1+q)^{-{2\over n}}\ .
\ea

Accordingly, the transfer matrix in an LQC bounce is given by
\be
T=\LF
\begin{array}{cc}
1  &0 \\
-{q(2q+1)\sqrt{\frac{\pi }{\al}} \Gamma \left(-\frac{1}{2}-\frac{1}{n}\right)\over \Gamma \left(-\frac{1}{n}\right)[q(\al^{1\over n}(1+q)^{-{2\over n}}-2)-1]}&-1
\end{array}
\RF\ .
\ee
\section{Application II: Stability of the Emergent Cyclic Inflation Model}\label{sec:cyclic-inflation}
\subsection{Emergent Cyclic Inflation: a brief Review}
As an application of the superHubble tracking formula (\ref{suph-soln}), in this section we will look at the evolution of superHubble fluctuations in the emergent cyclic inflation (ECI) scenario,  a new  inflationary paradigm that has been  proposed in a series of articles~\cite{emergent-cyclic,AB-cyclic,BM-CI,BMS-cmb,BKM-exit,BKM-cyclic,DB}.
In the presence of a bounce mechanism,  ECI hopes to realize a geodesically complete non-singular inflationary space-time via a series of asymmetric short cycles. Our primary aim is to check whether such a cosmology can be stable. For an overview on the model and its  motivations see~\cite{essay}, but  here is the ECI paradigm  in brief:
\begin{itemize}
\item Cyclic Emergent Phase: The universe ``begins'' in a quasi-periodic phase of oscillation. By quasi-periodic we mean that in each cycle the universe grows a little more than it contracts. These oscillations become more and more symmetric and periodic as we go back  towards the past infinity~\cite{emergent-cyclic}. The cosmology is very similar to the ``emergent universe'' scenario advocated in~\cite{emergent1,emergent2}. The emergent phase, as envisioned in~\cite{emergent-cyclic}, works if the universe is closed and contains different species of matter which are not always in thermal equilibrium with each other so that in each cycle a little bit of entropy is produced giving rise to the slight asymmetry in the cycles. While this asymmetric growth is required for the universe to eventually be able to exit the cyclic emergent phase, if this asymmetry persisted all the way to past infinity, that would lead to a geodesically incomplete universe with the amplitude of scale factor fluctuations vanishing as $t\ra -\infty$. However, it was first argued in~\cite{emergent-cyclic} and later numerically verified in~\cite{DB} that, in the cyclic emergent model as one goes back in the past the entropy production becomes less and less in every cycle and the universe asymptotes to a periodic oscillatory phase as $t\ra -\infty$. Such a space-time geometry is obviously geodesically complete.

   In the cyclic emergent phase, the turnarounds occur when the negative curvature density associated with the  closed universe cancels the total matter density, while for the bounces to occur one needs a separate mechanism in place.
\item Cyclic Inflation Phase: In the ECI scenario presented in~\cite{BMS-cmb,DB,essay}, as the universe grows slowly in the cyclic emergent phase, there comes a moment when the {\it negative} potential energy~\footnote{String theory, the leading candidate for a consistent theory of quantum gravity, naturally predicts the existence of negative energy vacua, and it has been quite a challenge to find ways that may lead to positive vacuum energies in the string theory/supergravity framework~\cite{sugra-ds,Kachru:2003sx}.
In general, if one looks at the potential energy coming from all the moduli in any fundamental theory,  one would expect to have both negative and positive potential regions, possibly with several local minima's dispersed liberally. In String theory this picture is often referred to as the ``landscape''~\cite{flux-compactification}.} (of various scalar fields that may be present) becomes more important than the negative curvature density. At this point the universe transitions to a cyclic inflation phase where  the   turnarounds now occur when the matter density is canceled by the approximately constant negative potential energy. One can easily check that now in every cycle has the same time period, and the volume of the universe grows by the same factor. The latter occurs because the interaction between different particle species  tend to naturally increase the entropy of the universe by the same factor:
     \be
     {S_{n+1}\over S_n}=1+3\ka\ ,
     \ee
     where $S_n$ denotes the entropy of the ``$n$th'' cycle. Since the total entropy is proportional to the volume, this means that for small $\ka$, the scale-factor (at say the bounce point) in consecutive cycles increases by a factor $(1+\ka)$. Over many cycles the space-time resembles the inflationary growth~\cite{BM-CI}!
\item Graceful Exit: In~\cite{BKM-exit} it was shown that the Cyclic inflationary phase can end if scalar field dynamics can classically propel the universe from a negative to a positive potential energy region. While this depends on several factors, such as how steep is the transition potential from the negative to positive potential regions, and what is the phase of the universe when the transition begins to occur, etc., it was found in~\cite{BKM-exit} that for most ``natural'' model parameters,  it is much more likely for the universe to make the exit  than not.
\item ``Our'' Expanding Universe: After the graceful exit the potential energy becomes positive and therefore, after one {\it last} bounce, the universe cannot turnaround any more. It has no other choice but to continue to expand  monotonically  which is where presumably we now find ourselves in. No reheating is required in the ECI paradigm as radiation remains the dominant energy density component throughout the entire evolution.
\end{itemize}

To summarize, the ECI paradigm provides a nonsingular geodesically complete inflationary cosmology, does not require any reheating (which has it's own challenges~\cite{infl-rev}), and interestingly predicts distinctive signatures in the form of small oscillatory wiggles in the power spectrum~\cite{BMS-cmb}. With the possible exception of the monopole problem the ECI can address all the old cosmological puzzles as in standard inflation.
Last but not the least, the CI mechanism provides new avenues of constructing phenomenologically viable inflationary models where the ``initial'' potential energy is negative. Keeping this in mind, here we want to address some of the questions that is pertinent to ascertain the viability of the ECI models.
\subsection{The Homogeneity Problem in Emergent Phase}
A key problem that often plagues cyclic and bouncing models is whether the assumption of an approximately homogeneous and isotropic cosmological metric holds during the bounce? The concern is well justified as it is  known that both anisotropy and inhomogeneity increases as the universe contracts to high densities. The problem is more profound in a model like ECI where the universe has to undergo an infinite set of cycles before reaching our current phase of monotonic expansion, and therefore encounters an infinite number of Planckian density bounces. Moreover, in the CI phase, after the comoving perturbations  exits into the superHubble phase, the universe still performs several oscillations~\footnote{The number of oscillations depend on the parameters of the model but for certain favored parameter sets, it was found that the universe undergoes around $\cO(10^3)$ cycles after the modes that we see in the CMB enters the superHubble phase~\cite{BMS-cmb}.}. Do the spectrum and amplitude of the perturbations relevant for CMB change during this process? Let us try to answer some of these questions with the help of the formalism we have developed in this paper.

We will start with the emergent phase where the universe is undergoing approximately a periodic cyclic evolution. A simple model that provides such a periodic evolution is provided by a LQC mediated bounce in the presence of a negative cosmological constant and radiation. This is essentially an extension of the scenario discussed in section~\ref{sec:turnaround}. The LQC correction (\ref{lqc}) provides a bounce when $\rho=\rhob$, while the turnaround occurs when $\rho_r=\La$. This gives us a periodic cyclic evolution of the universe much like one envisions in any emergent cosmology~\footnote{In the emergent-cyclic model the cycles are actually not exactly symmetric. Due to entropy production, in each cycle the universe grows a little bit more than it contracts. This is important for the universe to be able to exit the cyclic-emergent phase to the cyclic inflation phase. This slight asymmetry however, should not affect the evolution of the perturbations in any significant way and therefore we have decided to exclude the entropy generating mechanism in our analysis. For more details on this please see~\cite{DB}.}. For numerical purposes, one again has to resort to the acceleration equation which reads
\be
{\ddot{a}\over a}=-{1\over  M_p^2}\LT\6(\rho+3p)-{\rho\over \rho_{\bullet}}(3p+2\rho)\RT
\label{RCL}
\ee
where $\rho$ and $p$ refers to the total energy density and pressure given by (\ref{hubble}) and (\ref{pressure}).

\begin{figure}[htbp]
\begin{center}
\includegraphics[width=0.40\textwidth,angle=0]{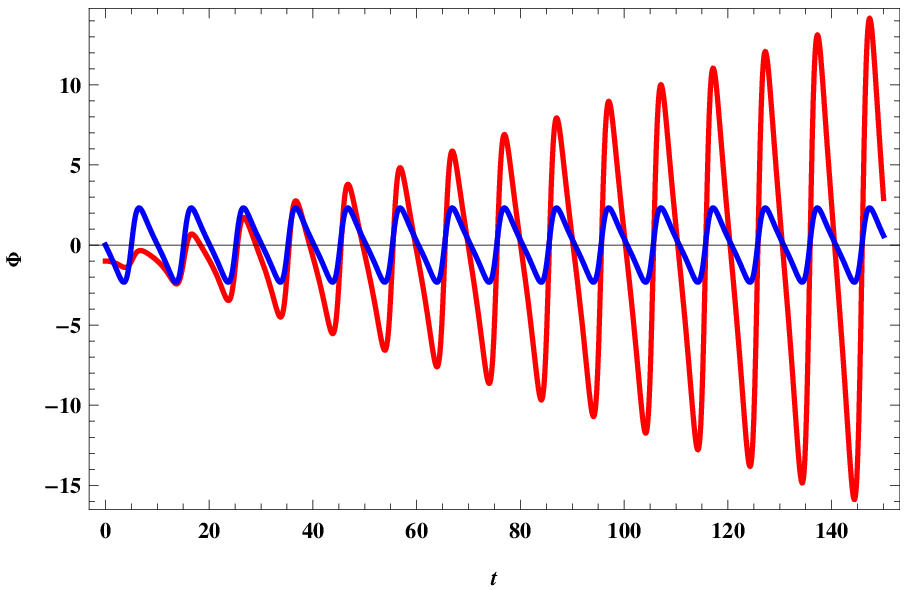}\\
\includegraphics[width=0.40\textwidth,angle=0]{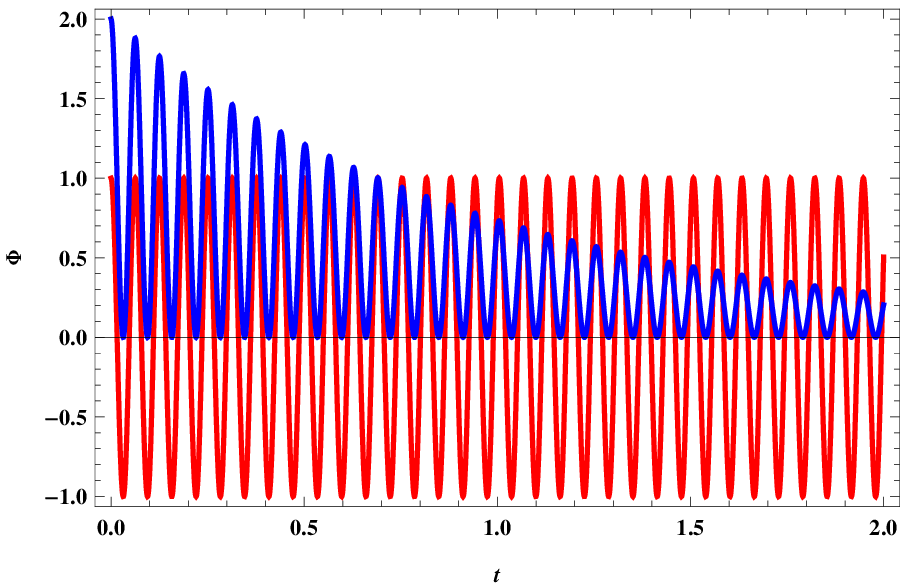}
\end{center}
\caption{The evolution of the $a_0$ mode (red) and the $b_0$ mode (blue) in the emergent cyclic (top) and the cyclic inflationary (bottom) phase.
  \label{fig:ECI}}
\end{figure}

(\ref{RCL}) can be solved straightforwardly, and then the two superHubble mode functions associated with $a_0,b_0$ can be plotted, see Fig.~\ref{fig:ECI} (top). The results are rather interesting: we observe that while the $b_0$ mode oscillates with constant amplitude, the $a_0$ mode starts to grow. This can be understood as follows: The expression for the $b_0$ model contains $\cH$ and $a$, both of which are periodic functions and therefore it is no surprise that the $b_0$ mode remains periodic. In other words, if we start in a ``patch'' of the universe where $b_0$ amplitude is small, it remains small during the emergent phase. The $a_0$ mode however, contains the function $g$ which is an integrand over a positive definite function and therefore can only increase. This results in the growth of the superHubble fluctuations in consonance with some earlier findings~\cite{cyclic-multiverse1,cyclic-multiverse2,Piao}.

This is a rather important and interesting result for the ECI model as it suggests that only patches which are smaller than the cosmological time scale of the bounce can remain smooth and homogeneous; fluctuations with wavelengths longer than this time scale becomes large. For a Planckian bounce, this means starting out with Planck size patches.

Addressing whether the amplitude of the subHubble (subPlanckian, in our case) fluctuations can remain small or not, is beyond the scope of our paper, but in a radiation dominated universe radiation pressure is expected to be able to prevent the growth of small scale fluctuations. In other words, Jean's instabilities are only expected to grow at wavelengths that are larger than the cosmological time scale. This is what we will also see in the following section in a scalar field driven bounce inflation model. Thus, a plausible picture that emerges in the ECI scenario is that of a ``patchy'' universe, where some of these Planckian patches can be homogeneous and isotropic~\footnote{The anisotropies are expected to grow as $\sim 1/a^6$. Since in the emergent cyclic phase the cycles essentially become periodic at past infinity, the scale factor at the bounce point asymptotes to a finite value. Unlike inhomogeneities, the anisotropies therefore don't grow and can remain under control.} that may eventually lead to a universe such as ours after a cyclic inflationary phase.
\subsection{SuperHubble Fluctuations in Cyclic Inflation Phase}
We have just witnessed that as long as the cyclic evolution is periodic the superHubble modes grow and therefore one can only envision a smooth isotropic patch of the size of the bounce time scale. We now want to ask what happens to the  superHubble perturbations if the universe starts to grow as envisioned in cyclic inflation. The detailed dynamics of the asymmetric cyclic growth depends on the specific model of cyclic inflation, but the advantage of the formalism that we have developed is that we only need to know the background cosmological evolution in order to track the long wavelength fluctuations. For the purpose of illutration, we will therefore look at a simple cyclic inflation background:
\be
a(t)=e^{\la t}[1+\en\sin(\oa t)]\with \en\ll1\mand \oa\gg \la\ .
\ee
In Fig.~\ref{fig:ECI} (bottom) we have plotted the two mode functions for the above cyclic inflation background. We observe that one of the mode decays while the other mode oscillates with a constant amplitude. This can be understood easily by inspecting (\ref{suph}): In the CI phase, $H$ is periodic but $a$ grows. This means that the $b_0$ mode now decays as $\sim 1/a$. The $a_0$ mode, however, contains an additional $g$ function whose  growth  $\propto a(t)$. This cancels the $1/a$ decay to lead to a constant amplitude!

This is a very important result: Firstly, this means that the superHubble fluctuations do not grow and hence once we start with a homogeneous and isotropic (Planckian) patch, the superHubble fluctuations won't spoil the underlying FLRW behavior. Secondly, the $a_0$ mode which ultimately becomes the constant mode in the final expanding background, seems to be able to retain its primordial amplitude of fluctuation through the course of the cyclic evolution. This is what was assumed in~\cite{BMS-cmb} in order to successfully explain the nearly scale-invariant spectrum in CMB, and we have now verified the conjecture. To summarize, our analysis gives credence to the ECI model with a twist that one can only assume a homogeneous and isotropic patch of the size of the Planck  length (or bounce scale) to  {\it begin} with, but that is a perfectly acceptable scenario.
\section{Application III: Bounce Inflation}\label{sec:bounce-inflation}
\subsection{Background Cosmology}
Our final study involves a simple realization of the ``Bounce Inflation'' scenario where the usual inflationary cosmology is preceded by a nonsingular bounce. The bounce provides past geodesic completeness to the inflationary space-time and envisions a cosmological evolution bereft of the Big Bang singularity and conceptual issues related to the beginning of time. Phenomenologically, it was pointed out recently~\cite{BM-bounce-inflation} that the bouncing phase may be able to account for the anomalous blue tilt that has been observed in the low-$l$ part of the CMB spectrum~\cite{WMAP9,Planck}. Thus a study of perturbations in such a model is interesting in it's own right, but our primary interest lies in checking how well the formalism we have discussed in tracking superHubble perturbations can be applied to this scenario and  whether we can develop any intuition when it comes to the subHubble fluctuations.

\begin{figure}[htbp]
\begin{center}
\includegraphics[width=0.40\textwidth,angle=0]{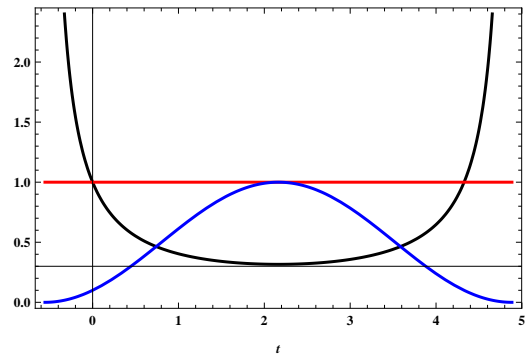}
\end{center}
\caption{Background Evolution in Bounce-inflation model. The black, red and blue curves represent $a(t)$, $\rho_{\phi}(t)$ and the absolute value of the curvature density, $M^4/a^2(t)$, respectively.
  \label{fig:bounce-inflation}}
\end{figure}

To this end let us consider a standard scalar field driven inflationary paradigm  evolving according to the usual Klein-Gordon equation:
\be
\ddot{\phi}+3H\dot{\phi}+{dV(\phi)\over d\phi}=0\ ,
\ee
which in conformal time becomes
\be
\phi''+2\cH\phi'+a^2{dV(\phi)\over d\phi}=0
\label{kg}\ .
\ee
The energy density and pressure of the scalar field are given by
\ba
\rho_{\phi}&=&\2\dot{\phi}^2+V(\phi)=\2\LF{\phi'\over a}\RF^2+V(\phi)\ ,\\
p_{\phi}&=&\2\dot{\phi}^2-V(\phi)=\2\LF{\phi'\over a}\RF^2-V(\phi)\ .
\ea
For the purpose of illustration, we are going to assume a simple slow-rolling potential,
\be
V(\phi)=M^4\LF v_0+v_1{\phi\over M_p}\RF,
\label{potential}
\ee
with $v_1\ll1$, which can drive a near exponential inflation. For a spatially flat universe the space-time, however, is geodesically incomplete. This problem can be avoided, if instead, we have a closed universe. In this case,  the total ``effective'' energy density and pressure that enters  the Hubble equation (\ref{hubble}) and the RC equation (\ref{rc}) are  given by
\ba
\rho&=&\rho_\phi-{M^4\over a^2}\\
p&=&p_{\phi}+\3 {M^4\over a^2}
\ea
The universe starts in a phase of contraction where for a sufficiently flat potential the scalar energy remains almost a constant while the negative curvature density increases in magnitude eventually canceling the scalar energy precipitating a bounce. After the bounce as the universe starts to inflate the curvature term vanishes, see Fig~\ref{fig:bounce-inflation}.
\subsection{Perturbations in a closed universe}
If the pre-bounce deflationary phase is allowed to continue all the way to past infinity, then all the perturbative modes would essentially start out superHubble. At some point during contraction they all become subHubble except the really long wavelength modes which remain longer than the bounce length scale even at the bounce. After the bounce, exactly the reverse happens, the modes which became subHubble starts to exit outside the Hubble radius as the universe inflates. In such a scenario it is not clear what initial conditions  one should choose for the various fluctuations as they start in the superHubble phase! For instance,  the all-important inflationary prediction of a near scale-invariant spectrum of perturbations that relies on being able to impose the Bunch Davis (BD) vacuum initial conditions in the subHubble phase, would no longer be applicable. Accordingly, phenomenologically a more interesting scenario would be to have a non-deflationary (say, matter or radiation dominated) contraction phase prior to deflation. In this case all modes would start out subHubble where the BD initial conditions can be imposed. Some of them will exit the Hubble radius during this phase. Once the deflationary phase begins though, all the modes that haven't exited will continue their journey as subHubble modes all the way to the expansion phase. It is possible therefore for these subHubble modes to exit the Hubble radius only once during the familiar inflationary phase, thereby reproducing the near-scale-invariance of the ensuing primordial spectrum. Can this process work?

Among the modes that do become superHubble during contraction, some of them re-enter the Hubble horizon during the deflationary phase, while others  remain superHubble forever. What kind of transformations do these perturbations undergo during the bounce? Clearly these are key questions for various model building exercises. Our aim in this paper is not to try and construct models, but rather to shed light on how perturbations (sub and super) evolve during the bounce which will aid for future model building. As we will see, our study will also provide us ways to go beyond the formalism and results we have discussed in the earlier sections.

In dealing with fluctuations from scalar fields, often it is convenient to work with the canonical variable
$\chi$ defined via (\ref{chi-defn})~\cite{Brandenberger} where
\ba
\cR&\equiv&\LF\Phi+H{\da \phi\over \dot{\phi}}\RF\non
z&\equiv& -a\sqrt{2\en_1[1+K_c\en_1/(k^2-4K_c)]^{-1}}\non
 K_c&=&{M^4\over 3M_p^2}\non
\mand \en_1&\equiv&{\dot{\phi}^2\over 2H^2M_p^2}={\phi^{'2}\over 2\cH^2M_p^2}\ .
\ea
Above, $\da\phi$ refers to the fluctuation of the scalar field.

In~\cite{closed-inflation} the evolution equation of $\chi$ was derived for a closed universe:
\be
\chi''_k+\LT k^2-3K_c-2{K_c z'\over \cH z}-{z''\over z}\RT\chi_k=0
\label{chi}
\ee
We observe is that at the bounce point when $\cH=0$, the fourth term in the evolution equation blows up. This is sometimes regarded as evidence of the fact that perturbations grow in an unbounded manner at the bounce leading one to argue that bouncing cosmologies are physically untenable or problematic. On the contrary, what this really indicates is simply that $\chi$ is  not a good variable to use at the bounce point! Instead if one uses, $\Phi,\da\phi$, to track the fluctuations one finds completely smooth transitions around the bounce, just as we saw around the turnaround cosmology in section~\ref{sec:turnaround}.

To demonstrate this, let us look at the evolution equations for $\Phi,\sa\equiv M_p\da\phi/\phi'$ in a closed universe~\cite{closed-inflation}:
\ba
\Phi_k'+\cH\Phi_k&=&{1\over 2M_p^2}\phi^{'2}\sa_k\\
\sa_k'+\cH\sa_k&=&\LT1+{2M_p^2(-k^2+3K_c)\over \phi^{'2}}\RT\Phi_k\ .
\label{closed-perturbations}
\ea
By inspection it is clear that the above equations remain regular during the bounce and can indeed be solved to yield perfectly regular solutions for $\Phi,\sa$. Fig.~\ref{fig:pert-closed}, top, middle and bottom represent the evolution of $\Phi,\sa$ and $\chi$ respectively. We notice that all the variables, including $\chi$ are finite and well defined. However $\chi$  is discontinuous at the bounce, and this is the reason why it cannot be tracked directly from its evolution equation (\ref{chi}). It's also obvious now that this does not represent any physical instability and one just has to use the appropriate perturbative variable to track the fluctuations across the bounce. In our opinion, this is quite an important result and along with our similar findings for the turnarounds should go a long way to put to rest some of the concerns regarding the consistency of bouncing cosmologies.
Let us now try to glean out some general patterns of evolution of both the sub and super Hubble fluctuations.
\begin{figure}[htbp]
\begin{center}
\includegraphics[width=0.40\textwidth,angle=0]{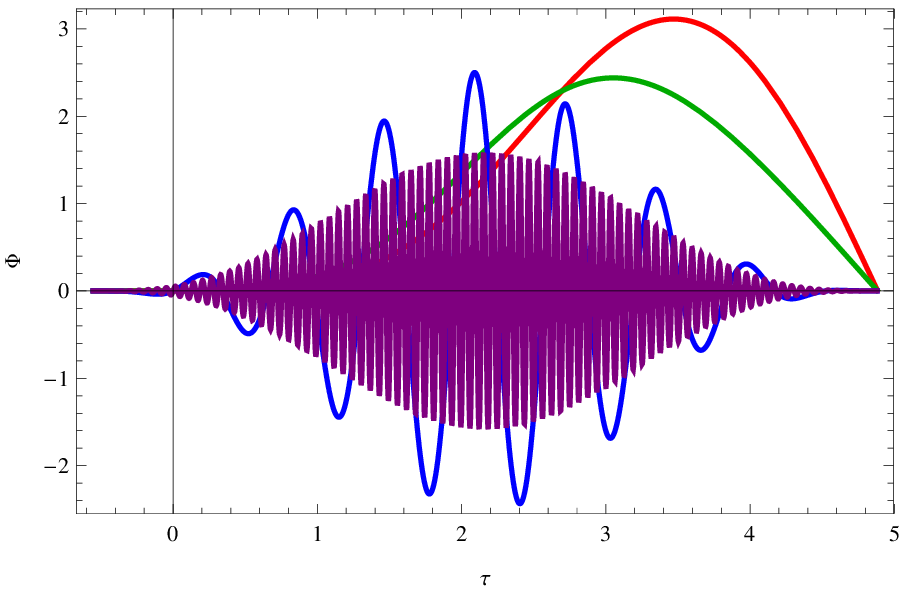}
\includegraphics[width=0.40\textwidth,angle=0]{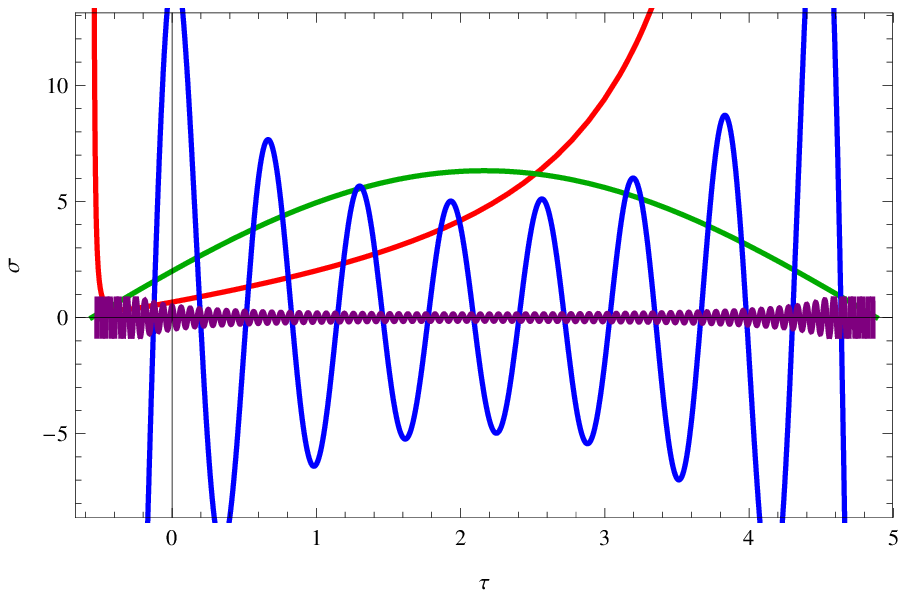}
\includegraphics[width=0.40\textwidth,angle=0]{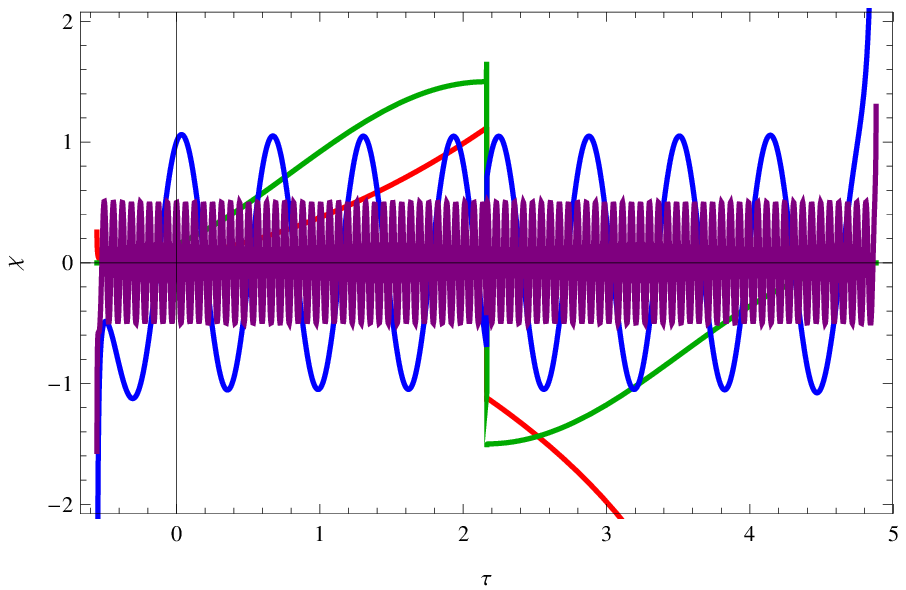}
\end{center}
\caption{Plots of $\Phi$ (top), $\sa$ (middle) and $\chi$ (bottom) for different comoving wavenumbers, $k=$ 0.01 (red), 1 (green), 10 (blue) and 100 (purple) for the case $M=M_p=1$. Our initial conditions are $a(0)=0.1$, $\phi(0)=1$ and $\phi'(0)=0$.
  \label{fig:pert-closed}}
\end{figure}
\subsection{Evolution of Fluctuations}
\subsubsection{SubHubble Evolution}
In Fig.~\ref{fig:pert-closed} we have plotted  $\Phi, \sa$ and $\chi$ as functions of time for different comoving fluctuations. As is evident from inspection, when the wavelength becomes smaller than the bounce scale (\ie subHubble) the amplitude of oscillations for the $\chi$ variable freeze. This is exactly what was found for the turnaround phase as well, see section~\ref{sec:turnaround}.  Indeed such a behavior is well known from analysis of subHubble perturbations in inflationary cosmology, but it is rather pleasing to see that the trend continues in more complex cosmologies lending credence to the fact that this may be a general feature of subHubble fluctuations.

What can one expect if and when one includes quantum gravitational effects that should be relevant for small wave-length fluctuations? SubHubble perturbations in FLRW backgrounds are actually very similar to fluctuations in Minkowski space-time as the wavelengths are too small to see the cosmological expansion. The only effect of the expansion is to stretch the wavelengths. This correspondence becomes particularly  visible in conformal time when one works with the ``canonical variable'', $\chi$. Barring near the bounce, it simply undergoes sinusoidal oscillations. In~\cite{BGKM,BKMV} a very general class of higher derivative theories of gravity, which may be able to incorporate several QG effects, have been studied. In particular it was shown that the perturbations break up into different modes, each mode effectively evolving according to its own Klein-Gordon type equation with a particular mass. This strongly suggests that in the subHubble regime we may continue to see simple sinusoidal oscillatory behavior in the appropriate canonical variables. In other words, we expect to only see an adiabatic stretching of the wavelengths of the subHubble fluctuations  with the evolution of the universe, while its amplitude remains a constant.

This would, for instance, strongly support the idea that even if the standard inflationary phase is preceded by a bounce, the $1/\sqrt{k}$ dependence of the ``initial'' amplitude of subHubble fluctuations, which comes from imposing the Bunch-Davis initial conditions in the contracting phase, will be preserved in the course of the bounce, thereby enabling the inflationary mechanism to provide the observed  near scale-invariant seed fluctuations for CMB, as usual.
\subsubsection{SuperHubble Evolution}
What happens to the superHubble fluctuations? The above model actually represents a case where neither of our superHubble formulas, (\ref{suph-soln}) and (\ref{suph-soln2}), are applicable. (\ref{suph-soln}) is not applicable because we have a closed universe where the scale-factor symmetry is no longer valid, while (\ref{suph-soln2}) works for a closed universe but only for ideal fluids; we have a scalar field. In fact, we especially chose this model to, on one hand, show the limitations of our superHubble formalism, but on the other, suggest how progress can still be made in spite of it. For such cases, one has to know the perturbative evolutions equations which can be solved numerically.
What is useful though is that our formalism can still help in obtaining the transfer matrix that connects the modes in contraction phase with modes during expansion. This is because we know that even in this case, once the curvature term becomes small, our superHubble formalism becomes applicable and therefore the perturbations should still behave according to (\ref{GRsup}) at late times. We thus expect that even in these models where our approach cannot directly be used to track the superHubble fluctuations during the entire cosmological evolution, with the help of numerics, we can still get a handle on the evolution of these fluctuations. Let us see this explicitly in our specific model now.

We know that at large times in the past and future, the spatial curvature is negligible and therefore the perturbation equation for $\Phi$ must reduce to the standard form~\cite{Riotto}:
\be
\Phi_k^{''} +2(\eta-\en) \cH\Phi_k^{'} + k^2\Phi_k +2(\eta-2\en)\cH^2\Phi_k \, = \, 0 \, ,
\ee
where the scalar field is assumed to be ``slowly rolling'', and
\ba
\en&=&{M_p^2\over 2}\LF{V'\over V}\RF^2\\
\eta&=&M_p^2\LF{V''\over V}\RF^2
\ea
are the  slow-roll parameters.
For our simple potential (\ref{potential}), as long as
\be
{\phi\over M_p}\ll {v_0\over v_1}
\ee
the slow-roll approximation should be valid with
\be
\en\approx  {v_1^2\over 2v_0^2}\mand \eta=0
\ee

For an exponential contraction or expansion the scale factor is given by
\be
a(\tau)={\mp}{1\over \la(\tau-\tau_{\pm})}\mand \cH=\mp{1\over \tau-\tau_{\pm}}\ ,
\ee
where $\tau\ra \tau_+$ from below for expansion and $\tau\ra \tau_-$ from above for contraction. In either case, the perturbation equation can be solved in the superHubble regime to yield
\be
\Phi=C_{\pm}(M_p\tau)^{q_1}+D_{\pm}(M_p\tau)^{q_2}
\ee
where,
\ba
q_1&=& -{4\en\over 1-2\en}\approx 0\\
q_2&=& \LT 1-2\en+{4\en\over 1-2\en}\RT\approx 1
\ea
In other words, approximately one can rewrite the superHubble modes as
\be
\Phi=C_{\pm}+{D_{\pm}\over a(\tau)}\ ,
\label{large-times}
\ee
precisely the form one obtains from our analytical formula (\ref{GRsup}).

\begin{figure}[htbp]
\begin{center}
\includegraphics[width=0.40\textwidth,angle=0]{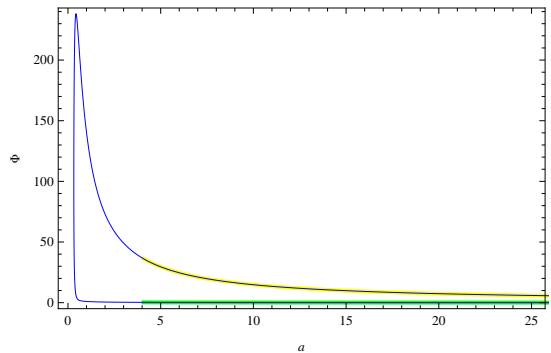}
\end{center}
\caption{The blue curve represents $\Phi(a)$ that we obtain numerically. The yellow and green curves represent the analytic approximations in the far past and distant future respectively. Our initial conditions are $a(0)=0.1$, $\phi(0)=1$ and $\phi'(0)=0$.
  \label{fig:pert-closed}}
\end{figure}
We can check this explicitly by solving the exact differential equations (\ref{closed-perturbations}) and then matching it with (\ref{large-times}) for large  times. To see this asymptotic  matching, it is more convenient to work with proper time as opposed to conformal time (the universe ends and begins abruptly in conformal time). In Fig.~\ref{fig:pert-closed}, we have plotted the evolution of $\Phi$ as the universe undergoes a bounce. As one can see, at large times, both in the past and future, it is fitted extremely well with (\ref{large-times}), represented by green and yellow curves respectively.

While fitting the asymptotic forms we could easily obtain the coefficients $C_{\pm},D_{\pm}$, and we noticed a somewhat surprising result. It seems that the constant mode amplitude actually changes during the course of the bounce as evident in Fig.~\ref{fig:pert-closed} as well. Also, the change in amplitude seems to vary  with $k$!
This means that the spectrum of the constant mode could be getting altered by contributions from the growing mode of the contraction phase. This is very different from the flat universe case where no such alteration was possible. This may be  good news for models based on ekpyrotic cosmology which naturally tend to produce a scale-invariant spectrum in the growing mode,  and warrants a more comprehensive investigation in the future. Our analysis can explains is why at times different results were obtained in different models when addressing the question of mode mixing: A first analysis of the
evolution of fluctuations through a specific nonsingular bounce
obtained by using a higher derivative gravity action was made in~\cite{Tsujikawa-flucts}, showing that initial
scale-invariant fluctuations do not pass through the bounce, thus
confirming the results of~\cite{Lyth,Robert-Finelli-ekpyrotic,Hwang,Khoury-Seiberg,Creminelli}. On the other hand, bounces
obtained by adding spatial curvature and matter with wrong-sign
kinetic terms, but with the standard gravitational action, have been
studied in~\cite{Finelli-curvature,Allen-curvature}, with differing results. It would also be most interesting to see  how our  analysis and results gets extended/changed when one ventures to more complicated cosmology involving many different degrees of freedom!
\section{Conclusions}\label{sec:conclusions}
Let us conclude by highlighting the main results of our paper and suggesting a few promising future research directions. We have established a comprehensive connection between the evolution of the superHubble fluctuations and background cosmology. This is manifested
 \begin{itemize}
 \item in our ability to obtain differential equations for the evolution of superHubble fluctuations from the evolution equations of background quantities like the scale-factor and various energy densities, as illustrated in the appendix
 \item in our ability to directly obtain the superHubble modes from the knowledge of all possible background solutions, and
 \item in our being able to derive explicit analytical formulas for two of the usual superHubble modes associated with the scale-factor and time-translation symmetry.
\end{itemize}
We were able to verify our results by explicitly solving perturbation equations analytically/numerically in some special toy models in GR where the perturbation equations were known. We not only looked at monotonically expanding and contracting universes but also more complex cosmologies involving bounces and turnarounds mediating nonsingular transitions between the contracting and expanding phases. This is particularly relevant for bouncing and cyclic cosmologies which were the main focus of the paper.

In the process,
\begin{itemize}
 \item we verified that perturbations can indeed remain small during the course of  bounces and turnarounds as long as one is careful to use appropriate variables,  divergences in ill-defined variables can indeed be misleading.
 \item We were able to prove that for a spatially flat universe, the constant mode in the expanding branch exclusively couples to the constant mode in the contracting branch confirming some of the earlier results. However in a closed universe the transfer matrix may be more interesting. This latter result needs to be investigated more comprehensively in the future due to it's obvious phenomenological implications.
 \item Our numerical forays  allowed us to corroborate the intuitive argument~\cite{Kinney,BMS-cmb} that although technically all modes are subHubble at the bounce/turnaround points, they behave as superHubble modes as long as their wavelengths are larger than the appropriate cosmological time scale controlled essentially by the energy density scale of the bounces and turnarounds in question.
 \item During our numerical analysis we found that the amplitude of the sinusoidal oscillations in the canonical perturbation variable, $\chi$,  remains a constant for the subHubble modes throughout the cosmological evolution. The fact that they should remain a constant in monotonically expanding and contracting phases was well known, but that it does so around the bounce and the turnaround is a pleasant surprise, especially considering that $\chi$ is ill-defined precisely at the bounce and turnaround. We provided further  arguments to suggest why this may be a very generic feature even when quantum gravitational effects are introduced.
 \item After testing our formalism in simple toy models, we applied our results to the emergent cyclic inflation scenario. In brief, our results indicated that superHubble fluctuations grow in emergent cyclic phase, so that one can only envision homogenous patches of Planck length. This homogeneity is however preserved in the cyclic inflation phase conforming the viability of this model.
 \item Next we applied our formalism to a bounce inflation model. We first checked that perturbations can remain finite and small during the bounce. Moreover, we found that the subHubble fluctuations can preserve its initial Bunch-Davis spectrum so that the generation of the near-scale-invariant spectrum of fluctuations in inflationary paradigm can operate in the usual way.
\end{itemize}

Throughout the course of this paper we have seen how our work and results can be generalized in different directions. Perhaps the most urgent and interesting venture would be to study a multi-component cosmological system and in particular study the possible generation of near scale-invariant perturbations in cosmologies such as new ekpyrotic, matter-bounce and hagedorn-bounce models. Another important direction would be to try to come up with a general framework to understand the subHubble fluctuations which, together with our results on superHubble fluctuations, would then provide a very concrete basis to construct and validate/reject bouncing/cyclic cosmological paradigms.

\section{Appendix}\label{sec:appendix}
In this appendix we illustrate how one may be able to derive the evolution equations that superHubble modes must satisfy from the knowledge of the background evolution equations by looking at the simple example of a single ideal fluid in GR. So, our starting point are the Eqns. (\ref{rc}) and (\ref{continuity}) describing the  evolution of the scale-factor and energy density in a homogeneous isotropic cosmological background. Now, small temporal fluctuations of the scale factor and the energy density is captured by
\ba
\cH&=&\cHb+\psi'\\
\mand \rho&=&\rb(1+\da)\ .
\ea
Then, perturbing the Hubble equation one obtains
$$2\cH \da\cH=\ka[2a\rho (\da a)+a^2(\da\rho)]$$
\be
\Ra 2\psi'=\cH(2\psi+\da)\ .
\label{pert-hubble}
\ee
 Similarly, perturbing the continuity equation one obtains
$$(\da\rho)'+3(1+\oa)[\rho(\da\cH)+\cH(\da\rho)]$$
$$=\da[\rho'+3(1+\oa)\cH\rho]+\rho[\da'+3(1+\oa)(\da\cH)]=0$$
where we have used the background continuity equation (\ref{continuity}). Finally we have
\be
\da'+3\psi'(1+\oa)=0
\label{pert-continuity}
\ee
(\ref{pert-hubble}) and (\ref{pert-continuity}) can now be combined to obtain a single differential equation for $\psi$:
\be
\psi''+\cH(1+3\oa)\psi'=0
\label{pert-psi}
\ee

We can now substitute $\psi$ in terms of $\Phi$ and $\xi$. For instance, we find
\ba
\psi'=-\Phi'-2\Phi\cH-\xi \cH'\nonumber\\
\psi''=-\Phi''-2\Phi'\cH--4\Phi\cH'-\xi \cH''
\label{derivatives}
\ea
Then using (\ref{relation}), and (\ref{derivatives})
We find
$$
\Phi''+3(1+\oa)\cH\Phi'+2[2\cH'+(1+3\oa)\cH^2]\Phi+\xi[\cH''+\cH\cH'(1+3\oa)]=0
$$
which reduces to the super-Hubble limit (\ref{super-hubble}) using the identity
\be
2\cH'+(1+3\oa)\cH^2=0
\label{identity}
\ee
valid in GR.

This demonstrates how starting from the homogeneous isotropic equations one can deduce the perturbation equation obeyed by $\Phi$ in the super-Hubble phase. A similar algorithm can be followed to find superHubble fluctuations from the background equations for any covariant theory of gravity. Note, that this procedure does not rely on symmetries such as the time translation or scale-factor symmetry and is therefore more general than the approach we have mostly followed in our paper. It should also be possible to generalize the algorithm to include scalar fields but we leave this as a future exercise.\vs
{\bf Acknowledgments:} TB would like to thank Abhay Ashtekar for initial collaboration and several helpful discussions especially on the subject of  Loop Quantum corrections to cosmological equations. TB would also like to thank Anupam Mazumdar and Robert Brandenberger for several insightful discussions on cosmological perturbation theory.
\bibliography{cyclicrefs}
\bibliographystyle{ieeetr}
\end{document}